\documentclass[%
reprint,
twocolumn,
superscriptaddress,
showpacs,preprintnumbers,
nofootinbib,
aps,
prd
]{revtex4-1}

\usepackage{subfigure}

\usepackage{amsmath}
\usepackage{amssymb}
\usepackage{mathtools}

\usepackage{hyperref}

\usepackage{graphicx}

\usepackage{color}

\usepackage{ulem}

\usepackage{comment}

\graphicspath{
  {./}
  {figures/ubar_plots/}
  {figures/fluidprofiles/}
  {figures/GW_plots_with_fits/}
  {figures/velocity_plots/}
  {figures/sensitivity/}
  {figures-erratum/}
}

\def\al{\alpha} 
\def\be{\beta} 
\def\ga{\gamma}
\def\de{\delta}
\def\ep{\epsilon}

\def\th{\theta}

\def\la{\lambda}

\def\si{\sigma}
\def\ta{\tau}

\def\om{\omega}
\def\De{\Delta}
\def\Ga{\Gamma}

\def\Om{\Omega}

\newcommand{\ben}{\begin{equation}}
\newcommand{\een}{\end{equation}}
\newcommand{\bea}{\begin{eqnarray}}
\newcommand{\eea}{\end{eqnarray}}
\newcommand{\ba}{\begin{array}}
\newcommand{\ea}{\end{array}}
\newcommand{\bit}{\begin{itemize}}
\newcommand{\eit}{\end{itemize}}
\newcommand{\vev}[1]{\left\langle#1\right\rangle}

\newcommand{\half}{\frac12}

\newcommand{\bk}{\textbf{k}}
\newcommand{\bq}{\textbf{q}}

\newcommand{\debar}[2]{{(2\pi)^{#1}}\de({#2})}

\newcommand{\Tfield}{\tau^\phi}
\newcommand{\Tfluid}{\tau^\text{f}}
\newcommand{\fluidT}{\tau_\text{f}}

\newcommand{\cs}{c_\text{s}} % Sound speed
\newcommand{\TN}{T_\text{n}} % Nucleation temperature
\newcommand{\tN}{t_\text{n}} % Nucleation time
\newcommand{\Tc}{T_\text{c}} % Critical temperature
\newcommand{\vw}{v_\text{w}} % Wall velocity
\newcommand{\vCJ}{v_\text{CJ}} % Chapman-Jouguet wall velocity
\newcommand{\fieldV}{\overline{U}_\phi} % Mean field gradient
\newcommand{\fluidV}{\overline{U}_\text{f}}  % Mean velocity
\newcommand{\fluidVmax}{\overline{U}_\text{f}^\text{max}}  % Mean velocity, max
\newcommand{\fluidVmaxperp}{\overline{U}_{\text{f},\perp}^\text{max}}  % Mean velocity, perp max
\newcommand{\Rc}{R_\text{c}} % Critical droplet radius
\newcommand{\Rbc}{R_*} % Mean droplet radius at collision
\newcommand{\Hc}{H_*} % Hubble rate at transition
\newcommand{\HN}{H_\text{n}} % Hubble rate at nucelation
\newcommand{\Nb}{N_\text{b}}

\newcommand{\Rfluid}{L_\text{f}}

\newcommand{\quadPar}{\gamma}
\newcommand{\cubPar}{A}
\newcommand{\relgamma}{W}
\newcommand{\strengthPar}[1]{\alpha_{#1}}
\newcommand{\StrParA}{\strengthPar{w}}
\newcommand{\StrParB}{\al_{\th}}

\newcommand{\PspecGW}{\tilde{\mathcal P}_{\text{gw}}}
\newcommand{\SpecDen}[1]{P_{#1}}
\newcommand{\SpecDenGW}{\tilde P_{\text{gw}}}

\newcommand{\rGW}{\rho_\text{gw}}
\newcommand{\OmGW}{\Omega_\text{gw}}
\newcommand{\OmGWnow}{\Omega_{\text{gw},0}}
\newcommand{\OmGWscaled}{\tilde\Omega_\text{gw}}

\newcommand{\uetcTen}{\Pi^2}

\newcommand{\IntSca}{\xi}

\newcommand{\CorLen}{\ell}

\newcommand{\phiAtMin}{\phi_\text{b}}

\newcommand{\vfft}{\tilde{v}}

\newcommand{\EOS}{\om} % Eq of state.  w already taken by enthalpy
\newcommand{\AdInd}{\Ga} %Adiabatic index = 1+\om
\newcommand{\fluidL}{L_\text{f}}

\newcommand{\tShock}{\ta_\text{sh}}
\newcommand{\tColl}{t_\text{pc}}

\newcommand{\Vol}{{\mathcal V}}

\newcommand{\zp}{z_\text{p}} % peak z = kR

\newcommand{\fp}{f_\text{p}} % peak GW frequency
\newcommand{\fpnow}{f_{\text{p},0}} % peak GW frequency

\newcommand{\tf}{t_\text{f}} % peak boundary area

\newcommand{\fitfun}{C}

\definecolor{newgreen}{RGB}{10,100,20}
\definecolor{purple}{rgb}{0.5,0,0.5}
\definecolor{BLUE}{rgb}{0,0,1}

\def\lsi{\raise0.3ex\hbox{$<$\kern-0.75em\raise-1.1ex\hbox{$\sim$}}}
\newcommand{\lsim}{\mathop{\lsi}}

\begin{document}
\newcommand{\Sussex}{\affiliation{
Department of Physics and Astronomy,
University of Sussex, Falmer, Brighton BN1 9QH,
U.K.}}

\newcommand{\HIPetc}{\affiliation{
Department of Physics and Helsinki Institute of Physics,
PL 64, % (Gustaf H\"{a}llstr\"{o}min katu 2),
FI-00014 University of Helsinki,
Finland
}}

\title{Shape of the acoustic gravitational wave power spectrum\\ from a first order phase transition}
\author{Mark Hindmarsh}
\email{m.b.hindmarsh@sussex.ac.uk}
\Sussex
\HIPetc
\author{Stephan J. Huber}
\email{s.huber@sussex.ac.uk}
\Sussex
\author{Kari Rummukainen}
\email{kari.rummukainen@helsinki.fi}
\HIPetc
\author{David J. Weir}
\email{david.weir@helsinki.fi}
\HIPetc

% Style to use for fixed dates:
% \date{April 20, 2017}

\date{April 16, 2020}

\begin{abstract}
  We present results from large-scale numerical simulations of a first
  order thermal phase transition in the early universe, in order to
  explore the shape of the acoustic gravitational wave and the
  velocity power spectra.  We compare the results with the predictions
  of the recently proposed sound shell model.  For the gravitational
  wave power spectrum, we find that the predicted $k^{-3}$ behaviour,
  where $k$ is the wavenumber, emerges clearly for detonations.  The
  power spectra from deflagrations show similar features, but exhibit
  a steeper high-$k$ decay and an extra feature not accounted for in
  the model.  There are two independent length scales: the mean bubble
  separation and the thickness of the sound shell around the expanding
  bubble of the low temperature phase.  It is the sound shell
  thickness which sets the position of the peak of the power spectrum.
  The low wavenumber behaviour of the velocity power spectrum is
  consistent with a causal $k^{3}$, except for the thinnest sound
  shell, where it is steeper.  We present parameters for a simple
  broken power law fit to the gravitational wave power spectrum for
  wall speeds well away from the speed of sound where this form can be
  usefully applied.  We examine the prospects for the detection,
  showing that a LISA-like mission has the sensitivity to detect a
  gravitational wave signal from sound waves with an RMS fluid
  velocity of about $0.05c$, produced from bubbles with a mean
  separation of about $10^{-2}$ of the Hubble radius.  The shape of
  the gravitational wave power spectrum depends on the bubble wall
  speed, and it may be possible to estimate the wall speed, and
  constrain other phase transition parameters, with an accurate
  measurement of a stochastic gravitational wave background.
 \end{abstract}

\preprint{HIP-2017-02/TH}

\maketitle

\section{Introduction}

The first direct observation of gravitational waves by the Laser
Interferometer Gravitational Wave Observatory (LIGO) in 2015 has
opened a new and unexplored window to the
cosmos~\cite{Abbott:2016blz,Abbott:2016nmj}. Even more excitingly,
while the original detection was related to an astrophysical process,
the merger of two black holes, gravitational waves will also allow us
to directly probe processes in the very early universe, such as
inflation~\cite{Bartolo:2016ami}, topological
defects~\cite{Sanidas:2012ee,Sanidas:2012tf,Aasi:2013vna,Blanco-Pillado:2013qja}
and first order phase transitions~\cite{Caprini:2015zlo}. To study
gravitational wave signals from these and other sources the Laser
Interferometer Space Antenna (LISA) is due to launch about a decade
from now~\cite{Audley:2017drz}. The LISA Pathfinder mission has
recently demonstrated the technological feasibility of such a
mission~\cite{Armano:2016bkm}.

LISA will have particular sensitivity in the millihertz frequency
range, making it an ideal instrument to observe gravitational wave
signals from phase transitions in the electroweak era, corresponding
to roughly 10 picoseconds after the big bang. First order phase
transitions proceed via the nucleation, expansion, collision and
merger of bubbles of the low temperature (broken) phase. They can
source gravitational waves in a number of ways.  Firstly,
gravitational radiation is produced by the collisions of the bubble
walls, where the scalar order parameter changes from the symmetric to
the broken phase.  If the phase transition occurred in vacuum, this
would be the only source of gravitational waves. The energy momentum
tensor of this source is well approximated by the envelope of a
configuration of infinitely thin
shells~\cite{Kosowsky:1991ua,Kosowsky:1992rz,Kamionkowski:1993fg};
this is known as the envelope approximation~\cite{Huber:2008hg}. It
leads to a characteristic high-wavenumber (UV) fall-off of the
spectrum proportional to $k^{-1}$, where $k$ is the wave number, which
has been recently confirmed by both numerical simulations
~\cite{Weir:2016tov} and analytic modelling \cite{Jinno:2016vai}.

If the scalar bubbles expand in a hot plasma, as expected to be the
case in the early universe, friction from the plasma will slow down
the walls, which after initial acceleration will expand with a
constant speed $v_w$. In this case most energy released by the
transition will be transferred from the scalar field into the
plasma. Only a tiny fraction, on the order of microphysics scale to
bubble radius at collision, will remain in the scalar field. So the
scalar field source is completely negligible for a thermal
cosmological phase transition\footnote{According to
  Ref.~\cite{Bodeker:2009qy} for very strong thermal transitions,
  friction from the plasma will not stop the bubbles from accelerating
  towards the speed of light (``runaway bubbles"). However, it is
  expected that additional friction from higher order corrections
  related to particle production will modify this
  result~\cite{Bodeker:2017cim}. So runaway bubbles are expected to
  correspond to very fast standard detonations.}.  The energy
transferred to the medium can either go to heat or fluid
motion. Numerical simulations show that the energy momentum tensor of
the fluid after bubble collisions corresponds to an ensemble of sound
waves.  These sound waves in turn are an efficient source of
gravitational
radiation~\cite{1986MNRAS.218..629H,Hindmarsh:2013xza,Hindmarsh:2015qta}.
An analytical model of the velocity perturbations in this acoustic
phase of the transition based on a picture of the acoustic phase
fluctuations as overlapping shells of sound waves has recently been
proposed \cite{Hindmarsh:2016lnk}, in which the UV fall-off of the
spectrum is roughly $k^{-3}$, so distinctly different from the
transition in the vacuum case.  This means that an observation of such
a gravitational wave signal with LISA could allow to distinguish
between the two cases, giving valuable information on the nature
of the transition. For very strong transitions it is expected that the
acoustic phase turns over into a turbulent
stage~\cite{Kosowsky:2001xp,Nicolis:2003tg,Caprini:2006jb,Gogoberidze:2007an,Caprini:2009yp,Kahniashvili:2009mf,Kahniashvili:2012vt,Kisslinger:2015hua}. This
turbulence will continue to produce gravitational radiation until it
decays.

How do the different possible components of gravitational radiation
from a thermal first order phase transition compare?  Collisions of
the bubble walls contribute on the order of $\Rc/\Rbc$ which for
electroweak bubbles is about $10^{-14}$ and therefore completely
negligible.

The relative weight of the contributions from the acoustic phase and
the subsequent turbulent phase crucially depend on the strength of the
transition. For very strong transitions, the plasma will quickly enter
the turbulent phase, which will then have a noticeable if not
dominating impact on the resulting gravitational wave signal.

Sufficiently weak transitions will not become turbulent before
gravitational wave production is effectively switched off by cosmic
expansion. The spectrum will then be dominated by gravitational
radiation from sound waves. This will be discussed in more detail in
Section~\ref{sec:gravitationalwaves}.

Magnetic fields may be present at the electroweak phase transition,
generated either earlier in the history of the Universe
\cite{Durrer:2013pga}, or by charge separation at the bubble walls if
CP violation is present \cite{Baym:1995fk}.  In this case, any
turbulent flow will redistribute energy between the fluid and the
magnetic field towards equipartition \cite{Kahniashvili:2012uj}.
Magnetohydrodynamic turbulence is expected to have its own
characteristic gravitational wave signal
\cite{Caprini:2006jb,Kahniashvili:2009mf}, although there is still
significant uncertainty about the shape of the power spectrum.
  
The peak frequencies of these different comments are typically quite
similar. For example, according to Ref.~\cite{Caprini:2015zlo} the
peak frequency of gravitational waves obtained from
magnetohydrodynamics (MHD) is about a factor of 1.5 higher than that
from sound waves, but given our limited understanding of turbulence
this may change as a result of future research. Also this difference
may well depend on the parameters of the phase transition, such as the
wall velocity.

In this paper we present results from extensive numerical simulations,
building on earlier work reported in
Refs.~\cite{Hindmarsh:2013xza,Hindmarsh:2015qta}. We model the system
by a scalar order parameter field coupled to a relativistic fluid by
means of a phenomenological friction term. The resulting Klein-Gordon
equation coupled to relativistic hydrodynamics is solved on a
lattice.  We use it to study the acoustic phase at an unprecedented
level of accuracy.

We show detailed velocity and gravitational wave power spectra, for
both deflagrations and detonations, and compare them to the
predictions of the sound shell model. The UV power laws agree with the
model in the case of detonations, and the prediction of an
intermediate $k^1$ power law for wall speeds close to the speed of
sound is also corroborated.  We establish that there are two length
scales in the power spectrum: the mean bubble separation, and the
width of the sound pulse around the expanding bubble wall, the
``shell'' of the sound shell model.  The UV power law for
deflagrations is steeper than the model prediction, with an
interesting break or knee.

We show that the gravitational wave power spectrum for wall speeds
well away from the speed of sound can be modelled with a broken power
law and an amplitude proportional to the fourth power of the RMS fluid
velocity and to the ratio of the fluid flow length scale to the Hubble
length. We use the model to forecast the sensitivity of LISA
\cite{Audley:2017drz} to acoustically generated gravitational waves.

Characterising the fluid flow length scale by the mean bubble
separation, we find that the peak sensitivity is to transitions with a
mean bubble separation of order $10^{-2}$ of the Hubble length at a
transition with critical temperature $10^2\;\textrm{GeV}$.
Transitions generating an RMS fluid velocity of about $0.05$ (in
natural units) give rise to acoustic gravitational waves with
signal-to-noise ratio of about 10.

We also estimate the timescale on which the acoustic waves become
shocked due to non-linear evolution, which would cause the velocity
and gravitational wave power spectra to deviate from their acoustic
form. A significant part of the parameter space generating observable
gravitational waves is likely to feature shocks and eventually
turbulence, for which further simulation is required to establish an
accurate power spectrum.

When interfaced with a particular microphysics realisation of a phase
transition to provide the main input parameters, latent heat, bubble
size and wall velocity, our results allow accurate estimations of the
resulting gravitational wave signal. Conversely, the differences in
the shapes we observe point the way towards estimating the wall speed
and constraining combinations of other phase transition parameters
from accurate observations of a primordial gravitational wave power
spectrum.

In the following section we recap the physics of the acoustic
generation of gravitational waves after a first-order thermal phase
transition; in Section~\ref{sec:methods} we discuss our numerical
methods, highlighting aspects of our approach which differ from
Refs.~\cite{Hindmarsh:2013xza,Hindmarsh:2015qta}; our results for the
fluid velocity power spectrum can be found in
Section~\ref{sec:fluidvelocity} and for gravitational waves in
Section~\ref{sec:gws}. We then compare these results to the power law
ansatz used for the LISA Cosmology Working Group report
(Ref.~\cite{Caprini:2015zlo}) in Section~\ref{sec:lisa}. Our
conclusions are in Section~\ref{sec:conclusions}.

\section{Acoustic gravitational waves}

The source of the gravitational waves is shear stress in the system,
induced by the nucleation, explosive growth and merger of bubbles of
the Higgs phase.  These perturbations take the form of compression and
rarefaction waves laid down around the growing bubbles - that is, the
sound of the Higgs explosions.

\subsection{Thermodynamics}

The sources of shear stress are the order parameter $\phi$ and the
relativistic fluid to which it is coupled.  Because we need only the
transverse-traceless part of the energy-momentum tensor, it is
sufficient to consider as a source tensor $\tau_{ij} = \Tfield_{ij} +
\Tfluid_{ij}$, which is decomposed into fluid and field pieces
according to
\begin{equation}
\label{e:TauDef}
\Tfield_{ij} = \partial_i \phi \partial_j \phi, \quad\Tfluid_{ij} = \relgamma^2 wV_i V_j,
\end{equation}
where $w = \epsilon + p$ is the enthalpy density, $\ep$ is the energy
density, $p$ is the pressure, $V_i$ is the fluid 3-velocity, and
$\relgamma$ is the corresponding Lorentz factor.  Unless the
transition is strongly supercooled, most of the available energy of
the transition goes into thermal and kinetic energy of the fluid; the
scalar contribution is negligible.

It is useful to describe the overall amplitude of the fluid shear
stress by a root mean square (RMS) four-velocity $\fluidV$ defined
through
\begin{equation}
\label{e:fluidVdef}
\fluidV^2 =  \frac{1}{\bar w \Vol}\int_{\Vol} d^3x \, \Tfluid_{ii}, 
\end{equation}
where $\Vol$ is the averaging volume and $\bar w$ is the volume
averaged enthalpy density.  One can define a similar quantity for the
scalar field
\begin{equation}
\label{e:fieldVdef}
\fieldV^2 =  \frac{1}{\bar w \Vol}\int_{\Vol} d^3x \, \Tfield_{ii}.
\end{equation}
Although these are not quite the magnitudes of the transverse
traceless part of the shear stress, they are easy to compute, and do
have a direct connection to the gravitational wave amplitude for
random fields, as we shall see in Eq.~(\ref{e:GWPowSpe}).

The fluid energy density and pressure have a contribution from the
scalar order parameter of the phase transition $\phi$, through its
effective potential $V(\phi,T)$,
\begin{align}
  p(T,\phi) &= \frac{\pi^2}{90} g_* T^4 - V(\phi,T)\\
\epsilon(T,\phi) &= \frac{\pi^2}{30} g_*  T^4 + V(\phi,T) - T\frac{\partial V}{\partial T},
\end{align}
where $g_*$ is the effective number of relativistic degrees of
freedom.

Following \cite{Enqvist:1991xw,Ignatius:1993qn}, we use a simple
quartic form for the potential:
\begin{equation}
\label{e:ScaPot}
V(\phi, T) = \frac{1}{2} \gamma (T^2-T_0^2) \phi^2 - \frac{1}{3} \cubPar T \phi^3 + \frac{1}{4}\lambda\phi^4.
\end{equation}
The detailed form is not important: its function is to supply a
metastable state with a latent heat
\begin{equation}
{\cal L}(T) = w(T,0) - w(T,\phiAtMin)
\end{equation}
where $\phiAtMin$ is the equilibrium value of the field in the
symmetry-broken phase at temperature $T$.  The strength of the
transition can be parametrised by the ratio of the latent heat to the
total radiation density in the high temperature symmetric
phase\footnote{Note that the radiation density has no unique
  definition in a system with a scalar order parameter in a thermal
  bath: here we follow Ref.~\cite{Espinosa:2010hh} and define it as
  $\ep_\text{r}(T) = 3w(T,0)/4$.}
\begin{equation}
\StrParA = \frac{{\cal L}(T)}{\ep_\text{r}(T) }.
\label{e:StrParA}
\end{equation}
A commonly used alternative is the difference in the trace anomaly
divided by a conventional factor of 4, or
\begin{equation}
\De\th(T) =  - \frac{T}{4} \frac{d}{dT}\De V + \De V,
\end{equation}
where $\De V(T) = V(0,T) - V(\phiAtMin,T) $.  Expressed relative to
the energy density in the symmetric phase
\begin{equation}
\StrParB = \frac{\De\th(T)}{\ep_\text{r}(T)}.
\label{e:StrParB}
\end{equation}
Other important parameters are the surface tension and width of the
phase boundary $\si$ and $\CorLen$, which can be computed
straightforwardly from the parameters of the potential
\cite{Enqvist:1991xw,Hindmarsh:2015qta}.  The transition takes place
at the nucleation temperature $\TN$, at which the radius of the
critical bubble is denoted $\Rc$.

\subsection{Shear stress and velocity correlations}

As mentioned above, the dominant source of shear stress is the fluid,
unless $\al \gg 1$ by the measures of the phase transition strength
outlined above, and the scalar field is so weakly coupled with the
fluid that the walls continue to accelerate until collision.  Our
simulations explore the more generic situation, $\al \lsim 1$.

One can characterise the fluid source by the unequal time correlator
(UETC) of the shear stress $\uetcTen$
\cite{Caprini:2009fx,Figueroa:2012kw}, defined by projecting out the
spatially transverse and traceless part of the energy-momentum tensor
\begin{multline}
\la_{ij,kl}(\bk)\vev{\fluidT^{ij}(\bk,t_1)\fluidT^{kl}(\bk',t_2)} \\
= \uetcTen(k,t_1,t_2) \debar3{\bk + \bk'},
\label{e:UETCdef}
\end{multline}
where
\begin{equation}
\label{e:ProDef}
\la_{ij,kl}(\bk) = P_{ik}(\bk) P_{jl}(\bk) - \half P_{ij}(\bk)P_{kl}(\bk) 
\end{equation}
and
\begin{equation}
P_{ij}(\bk)  = \de_{ij} - \hat{k}_i \hat{k}_j. 
\end{equation}
The shear stresses are the result of the sound waves, which can be
characterised by the longitudinal part of the velocity unequal time
correlator $G(q,t_1,t_2)$.  This is defined from the Fourier transform
of the velocity field $\vfft^i_{\bq_1}$ through
\begin{equation}
\hat{q}^i \hat{q}^j \vev{\vfft^i_{\bq_1}(t_1)\vfft^{*j}_{\bq_2}(t_2)} = G(q,t_1,t_2) \debar3{\bq_1 - \bq_2}.
\end{equation}
The transverse part $G^\perp$ can be defined analogously.  We will be
interested in the velocity power spectrum,
\begin{equation}
\label{e:VPowSpe}
\frac{dV^2}{d \ln(q)} = \frac{q^3}{2\pi^2} \left(G(q,t,t) + G^\perp(q,t,t)\right).
\end{equation}
In our simulations, the longitudinal part is always much greater than
the transverse part (see Table \ref{t:SimVelStats}), reflecting the
dominance of sound waves in the fluid perturbations.

\subsection{Gravitational waves}
\label{sec:gravitationalwaves}

The transverse traceless metric perturbation $h_{ij}$ is extracted by
projection from an auxiliary tensor $u_{ij}$ satisfying the equation
\begin{equation} 
\Box u_{ij} = (16\pi G) \Tfluid_{ij}.
\end{equation}
The energy density in gravitational waves is
\begin{equation}
\rGW = \frac{1}{32\pi G}\left< \dot h_{ij} \dot h_{ij} \right>,
\end{equation}
where $h_{ij}$ is the transverse-traceless projection of $u_{ij}$. It
is most useful to consider the gravitational wave power spectrum
relative to the critical density, defined as
\begin{equation}
  \frac{d \OmGW}{d \ln(k)} = \frac{1}{12 H^2} \frac{k^3}{2\pi^2}
  \SpecDen{\dot h}(\bk,t),
\label{e:OmGWPowSpeDef}
\end{equation}
where $H$ is the Hubble parameter and $\SpecDen{\dot h}$ is the power
spectral density, defined from the two-point correlation
\begin{equation}
\vev{\dot{h}_{ij}(\bk,t) \dot{h}_{ij}(\bk',t) } = \SpecDen{\dot h}(\bk,t) \debar3{\bk+\bk'}.
\end{equation}
In our simulations, the expansion of the universe is scaled out using
the scale invariance of the relativistic fluid equations
\cite{Brandenburg:1996fc,Hindmarsh:2015qta}. Nonetheless, one can
still formally compute the Hubble rate from the Friedmann equation
\begin{equation}
  \label{eq:friedmanneq}
  H^2 = \frac{8\pi G}{3} \ep(0,T).
\end{equation}

It was shown in Ref.~\cite{Hindmarsh:2015qta} that, after the acoustic
source has been on for time $t$, the dimensionless gravitational wave
power spectrum takes the form
\begin{multline}
\label{e:GWPowSpe}
\frac{d \OmGW(k)}{d \ln(k)} \\ =  3\AdInd^2 \fluidV^4  (\HN t) (\HN\Rfluid) \frac{(k\Rfluid)^3}{2\pi^2} \SpecDenGW(k\Rfluid),
\end{multline}
where $\AdInd = 1 + \bar p/\bar \ep$ is the adiabatic index, $\fluidV$
is the RMS fluid velocity, $\HN$ is the Hubble rate at the bubble
nucleation temperature $\TN$, $\Rfluid$ is the characteristic length
of the fluid flow, and $\SpecDenGW$ is a dimensionless spectral
density for the gravitational waves.

It was also shown that, provided that turbulence does not develop
within a Hubble time, the effective time for which the acoustic source
operates is precisely the Hubble time, so that $\HN t \to 1$.

The turbulence timescale, both for appearance and decay, is the shock
appearance or eddy turn-over time~\cite{LanLifFlu,Pen:2015qta}
\begin{equation}
\tShock \sim \fluidL/\fluidV.
\end{equation}
The maximum duration of all our simulations is much less that
$\tShock$, and so no turbulence develops.  Our results therefore apply
to flows for which $\fluidV \ll \fluidL \HN $.

With this assumption, 
the total gravitational wave energy density
from the acoustic phase is
\begin{equation}
\label{e:GWTotPow}
\OmGW^\text{ac} = 3\AdInd^2 \fluidV^4  (\HN\Rfluid) \OmGWscaled,
\end{equation}
where
\begin{equation}
\label{e:OmTilDef}
\OmGWscaled = \frac{1}{2\pi^2} \int_0^\infty {dx} \, x^2 \SpecDenGW(x) 
\end{equation}
is a dimensionless parameter, quantifying the efficiency with which
shear stress is converted to gravitational waves.  A significant
result in \cite{Hindmarsh:2015qta} was that this parameter is
approximately independent of the length scale and RMS velocity of the
fluid flow.

The formula (\ref{e:GWPowSpe}) and (\ref{e:GWTotPow}) are derived from
general considerations of the velocity correlation function, so one
can also estimate the relative amplitude of gravitational waves
produced during the turbulent phase.  Eq.~(\ref{e:GWTotPow}) arises
from a flow with RMS velocity $\fluidV$ with characteristic length
scale $\fluidL$ starting when the Hubble parameter is $\HN$.  The
turbulent phase starts when the Hubble parameter is of order
$\tShock^{-1}$, with similar RMS velocity and characteristic length
scale.  Hence once can estimate that the density parameter of
gravitational waves produced during the turbulent phase is
\begin{equation}
\OmGW^\text{tu} \sim \AdInd^2 \fluidV^4  (\Rfluid/\tShock),
\end{equation}
a factor $(\Hc\tShock)^{-1}$ smaller than from the acoustic phase.
Therefore, for transitions with $\fluidV \ll \fluidL \HN $, the
gravitational wave signal from the turbulent phase can be
neglected. One reaches the same conclusion with a more careful
derivation based on Eq.\ (A9) of Ref.~\cite{Hindmarsh:2015qta}.

In the case of a strong phase transition, turbulence develops in less than a Hubble time, 
the lifetime of both the acoustic and turbulent phases is of order $\tShock$.
Hence, using $t \sim \tShock \sim \fluidL/\fluidV$ in (\ref{e:GWPowSpe}) 
and integrating over wavenumber, one finds 
\begin{equation}
\OmGW^\text{ac} \sim \OmGW^\text{tu} \sim \AdInd^2 \fluidV^3  (\Rfluid\HN)^2.
\end{equation}
Hence for strong phase transitions the acoustic and turbulent signals
should have similar magnitudes.

\section{Methods}
\label{sec:methods}

The system is a set of coupled partial differential equations
governing the evolution of the scalar field $\phi$ and the
relativistic ideal fluid with 4-velocity $U^\mu$. We use the
techniques previously described in \cite{Hindmarsh:2015qta} (see also
Refs.~\cite{KurkiSuonio:1995pp,KurkiSuonio:1995vy}, and the
textbooks~\cite{WilsonMatthews,RezzollaZanotti}).

The field and fluid parts of the system are coupled together through a
dissipative term that turns field stress-energy $T^{\mu\nu}_\phi$ into
fluid stress-energy $T^{\mu\nu}_\text{f}$ such that,
\begin{equation}
  \label{eq:couplingterm}
  \partial_\mu T^{\mu\nu}_\phi = - \eta U^\mu \partial_\mu\phi \partial^\nu \phi
\end{equation}
where $\eta$ is in general a function of $\phi$ and $T$ with mass
dimension 1.  In previous work, including our own
\cite{Hindmarsh:2013xza,Hindmarsh:2015qta}, this was taken to be a
constant.  In this work we take $\eta = \tilde\eta {\phi^2}/{T}$,
where $\tilde\eta$ is a dimensionless parameter, which is better
motivated by the underlying physics \cite{Moore:1995si,Huber:2013kj}.
The fluid velocity around the expanding bubbles for a given wall speed
and phase transition strength is minimally affected by the change, as
it is determined purely by hydrodynamics, except right at the bubble
wall \cite{Steinhardt:1981ct,KurkiSuonio:1995vy,Espinosa:2010hh}.

\begin{table}

\begin{tabular}{ c  c  c }
Parameter & Weak & Intermediate \\
\hline
$g_*$			& $106.75$	& $106.75$ \\
$T_0/\Tc$ 			&  $1/\sqrt{2}$ 	& $1/\sqrt{2}$ 	\\
$\quadPar$ 		& $2/9$ 		& $4/9$ 		\\
$\cubPar$ 		& $0.1990$ 	& $0.1990$ \\
$\la$ 			& $0.0792$ 	& $0.0396$ 	\\
$\TN/\Tc$ 			& $0.86$ 		& $0.80$		\\
\hline
${\cal L}/T_c^4$      	& $0.7013$ 	& $5.6102$ 		\\
$\sigma/T_c^3$ 	& $0.1558$ 	& $0.8816$ \\
$\CorLen \Tc$ 		& $3$ 		& $2.1213$ 	\\
\hline
$\phiAtMin(\TN)/\Tc$ 		& $1.7838$ 		& $3.5810$ 	\\
$\StrParA(\TN)$ 		& $0.010$ 	& $0.084$ 	\\
$\StrParB(\TN)$ 		& $0.0046$ 	& $0.050$ 	\\
$\vCJ$				& $0.63$			& $0.73$			\\
$\Rc \Tc$ 			& $8.1$ 		& $4.3$ 		\\
$ H_\text{n}/\sqrt{G} \Tc^2 $		& 	$12.686 $	& 	$10.978$	\\
\hline
\end{tabular}

\caption{Input parameters and derived equilibrium and non-equilibrium
  quantities for our simulations. Our parameters are the effective
  number of relativistic degrees of freedom $g_*$, scalar potential
  parameters [see Eq.~(\ref{e:ScaPot})] and nucleation temperature
  $\TN$. From these we obtain the latent heat ${\cal L}$, phase
  boundary tension $\si$ and the thickness $\CorLen$. For studying
  phase transitions, it is useful to also compute the equilibrium
  value of the scalar field at the nucleation temperature
  $\phiAtMin(\TN)$, transition strength parameters $\StrParA$ and
  $\StrParB$ [see Eqs.~(\ref{e:StrParA}) and (\ref{e:StrParB})], the
  Chapman-Jouguet speed for detonations $\vCJ$, and the critical
  bubble radii $\Rc$.  Finally, we use Eq.~(\ref{eq:friedmanneq}) to
  compute a value for the Hubble constant $H_\text{n}$.  }

\label{t:SimParsPot} 
\end{table}

The parameters chosen in the numerical simulations are given in Table
\ref{t:SimParsPot}.  All dimensionful quantities are expressed in
terms of the critical temperature $\Tc$, defined from $\De V(\Tc) =
0$.

We take the effective number of relativistic degrees of freedom to be
the Standard Model's high temperature value, although its exact value
is not important\footnote{From Eq.~(\ref{e:StrParA}) and the potential
  Eq.~(\ref{e:ScaPot}), one can see that the phase transition strength
  $\StrParA$ can be kept constant if $g_*$, $\gamma$, $\cubPar$ and
  $\lambda$ are all scaled by some constant $C$. This will change the
  correlation length $\CorLen$ and hence the bubble wall thickness by
  a factor $\sqrt{C}$ but will otherwise have little impact on the
  position or amplitude of the resulting gravitational wave power
  spectrum, so long as Eq.~(\ref{eq:scaleseparation}) holds.}.  The
value of Newton's constant $G$ is arbitrary, as we will compute
quantities which are independent of $G$.

We simulate at two different transition strengths, which we label
``weak'' ($\al \sim 10^{-2}$) and ``intermediate'' ($\al \sim
10^{-1}$). These have the same phase transition strengths as our
``weak'' and ``intermediate'' simulations in
Refs.~\cite{Hindmarsh:2013xza,Hindmarsh:2015qta}, but the different
form of the field-fluid coupling term, and the changed value of the
relativistic degrees of freedom $g_*$ mean that the simulations are
not identical. The differences, however, are minor and do not affect
the results.

For both weak and intermediate categories, we simulate at a variety of
wall velocities giving both deflagrations and detonations, including
some which move at close to the Chapman-Jouguet speed $\vCJ$, defined
as the wall speed at which the exit velocity of the fluid in the wall
frame is the speed of sound \cite{LanLifFlu}.  For a deflagration,
$\vCJ = \cs$, while for a detonation it depends on the strength of the
transition, but is always greater than $\cs$ (see
e.g.~Ref.~\cite{Espinosa:2010hh}).  For a weak transition, $\vCJ
\simeq \cs (1 + \sqrt{2\StrParB} ) $.  Note that the values of the
transition strength parameter in Table~\ref{t:SimParsPot} do not take
into account the small increase in the temperature of the fluid near
the moving bubble wall, and hence the stated Chapman-Jouguet speed is
slightly higher than the true minimum speed of a detonation.

Each bubble is nucleated by inserting a scalar field configuration
with Gaussian profile, as described in Ref.~\cite{Hindmarsh:2015qta}.
This profile is slightly larger than the critical bubble radius $\Rc$.
The bubble expands and perturbs the fluid, which evolves towards the
scaling solution (c.f.  Fig.~\ref{fig:fluidprofiles}).
 
Our bubbles are nucleated simultaneously, rather than with a physical
nucleation rate. It is possible to rescale the results to yield the
power spectrum from a more realistic nucleation rate. This was
demonstrated in Ref.~\cite{Weir:2016tov} and is discussed further in
Section~\ref{sec:lisa}. We simulate with a variety of bubble numbers
$\Nb$, which controls the average bubble separation
\begin{equation}
  \Rbc = (\Vol/\Nb)^{\frac{1}{3}}.
\end{equation}
After time $t$, each bubble is surrounded by a sound shell
\cite{Hindmarsh:2016lnk} of approximate thickness $\De R = \De\vw t$,
where $\De \vw = |\vw - \cs|$.

For near-Jouguet bubble walls the sound shell is typically quite thin
relative to the bubble separation.  In the sound shell model
\cite{Hindmarsh:2016lnk}, the velocity power spectrum peaks at
$k\De\Rbc \sim 1$, where $\De\Rbc$ is the average bubble shell
thickness at collision.

The bubble wall speeds and bubble separations are chosen to explore
the dependence of the power spectra on $\Rbc$ and $\De\Rbc$, and also
to compensate for the limited dynamic range of the simulations.  The
values of $\Rbc$ and $\De\Rbc$ are listed in Table
\ref{t:SimParsRuns}.

In principle we need sufficiently large lattices to explore
\begin{equation}
\label{eq:scaleseparation}
dx \ll \CorLen \ll \De\Rbc \lesssim \Rbc \ll \fluidL.
\end{equation}
In practice, the bubble wall does not need high resolution, as the
detailed dynamics of the scalar field are not important beyond the
transfer of energy to the fluid.  For the near-Jouguet transitions, we
found it useful to explore the wavenumber range $k\Rbc \ll 1$ and
$k\De\Rbc \gg 1$ separately by adjusting the number of bubbles.
Simulations with large $\Rbc$ have the advantage that the velocity
field is closer to its asymptotic self-similar form.

Velocity and gravitational wave power spectra are computed on cubic
periodic lattices with $N=4200$ points per side.  The lattice spacing
$dx$, bubble number $\Nb$, and bubble wall speed $\vw$, along with the
field-fluid coupling required to obtain this speed, are listed in
Table \ref{t:SimParsRuns}.

We generally run with $dx=2/\Tc$ which has been established to work
well for single bubble self-collisions for a weak deflagration and
$\Rbc \approx 384/\Tc$~\cite{Hindmarsh:2015qta}.

Wall velocities $\vw$ depend on discretisation effects, and it is
difficult to determine the final asymptotic value from numerical
simulations of limited duration. For concreteness, our quoted $\vw$
values are determined from spherically symmetric simulations of a
single bubble with $dx\,\Tc=0.2$ and are measured at time
$5000/\Tc$. For coarser lattice spacings the actual wall velocity will
be slightly smaller, but the difference in the cases studied here is
at most 3\% (this effect is more pronounced for faster wall
velocities~\cite{CuttingThesis}).  In our simulations, the bubbles
collide before the asymptotic profile is reached (see
Fig.~\ref{fig:fluidprofiles}), and this is generally more significant
than discretisation effects, particularly for simulations with smaller
mean bubble separations.

For simulations with $\vw$ close to $\cs$ the lattice discretisation
effects on the fluid profile were more significant than for other
cases. In the weak, $\vw = 0.59$, $\Rbc \approx 1900/\Tc$ case, we ran
simulations at both $dx = 1/\Tc$ and $dx = 2/\Tc$ as a check against
lattice effects in our final results. Agreement in the velocity power
spectra is excellent until $k \Rbc \approx 100$, deteriorating to an
error of about $50\%$ at $k \Rbc \approx 200$. The discrepancy is more
pronounced in the gravitational wave power spectrum because it
convolves the power at different wavelengths~\cite{Hindmarsh:2016lnk}.

\begin{table}
\begin{tabular}{l c c c c c c}
  Type      & $\tilde\eta$ &  $\Nb$ & $dx\Tc$ &  $\vw$ & $R_*\Tc$ &  $\Delta R_*\Tc$\\
\hline
Weak 	& 0.19 & 84 & 2 &  $0.92$ & 1918 & 714.4\\
 		& 0.35 & 84 & 2 &  $0.80$ & 1918 & 533.8\\
 		& 0.51 & 84 & 2 &  $0.68$ & 1918 & 289.5 \\
 		& 0.59 & 11 & 1 &  $0.56$ & 1889 & 58.51\\
 		& 0.93 & 84 & 2 &  $0.44$ & 1918 & 598.7\\
\cline{2-7}
		& 0.51 & 5376 & 2 &  $0.68$ & 480 & 72.38\\
		& 0.59 & 5376 & 2 &  $0.56$ & 480 & 14.86\\
		& 0.93 & 5376 & 2 &  $0.44$ & 480 & 149.7\\
\hline
Int. 	& 0.17 & 84 & 2 &  0.92 & 1918 & 714.4\\
 		& 0.40 & 11 & 1 &  0.72 & 1889 & 374.2\\
 		& 0.62 & 84 & 2 &  0.44 & 1918 & 598.7\\
\hline
\end{tabular}
\caption{\label{t:SimParsRuns} Simulation parameters $\tilde\eta$
  (field-fluid coupling), $\Nb$ (number of bubbles nucleated) and
  lattice spacing $dx$, with the corresponding bubble wall speed
  $\vw$, mean bubble separation $\Rbc$ and sound shell width $\Delta
  \Rbc = \Rbc |\vw - \cs|/\cs$.  The potential parameters and derived
  quantities for the ``weak" and ``intermediate" phase transitions are
  given in Table~\ref{t:SimParsPot}.  }
\end{table}

\section{Results: fluid velocity}
\label{sec:fluidvelocity}

\begin{figure*}
  \subfigure[\ Weak, $\Rbc=1900/\Tc$]{\includegraphics[width=0.32\textwidth]{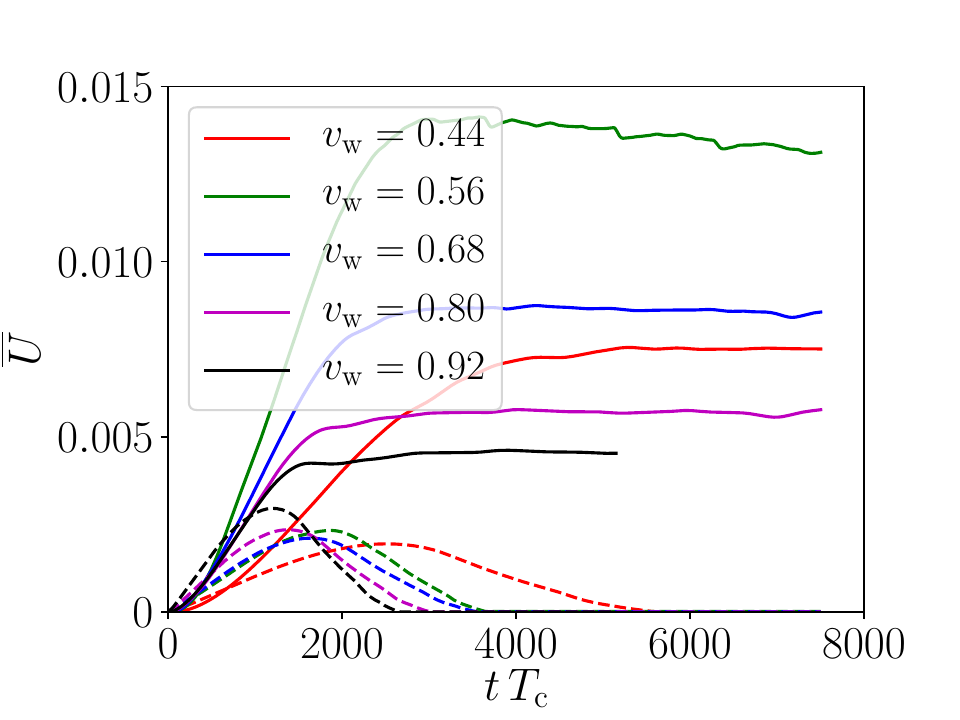}}
  \hfill
  \subfigure[\ Weak, $\Rbc=480/\Tc$]{\includegraphics[width=0.32\textwidth]{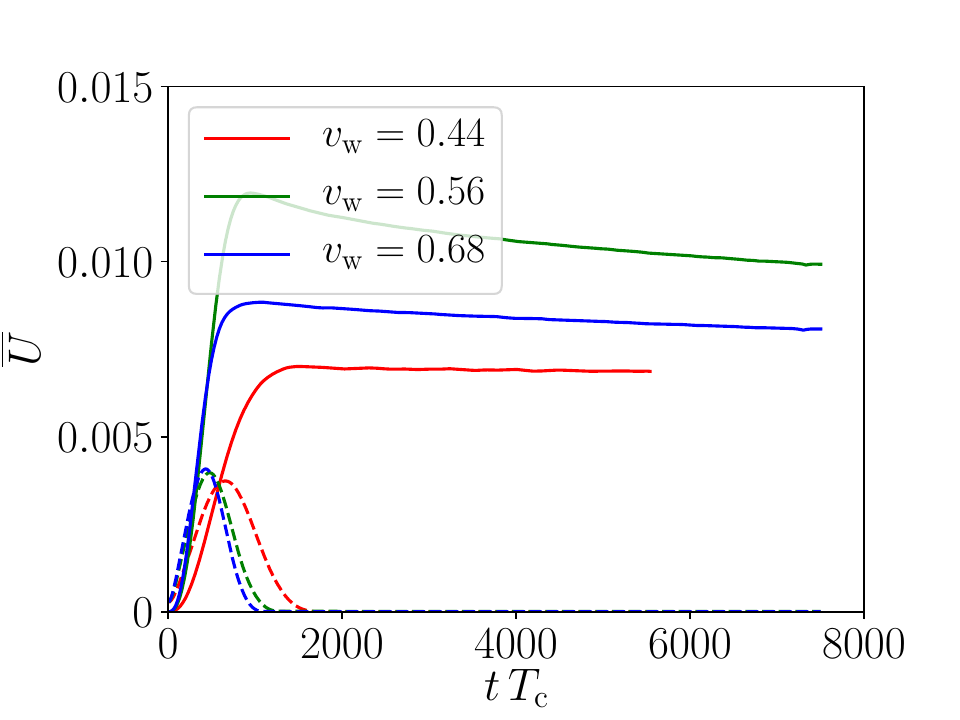}}
  \hfill
\subfigure[\ Intermediate, $\Rbc=1900/\Tc$]{\includegraphics[width=0.32\textwidth]{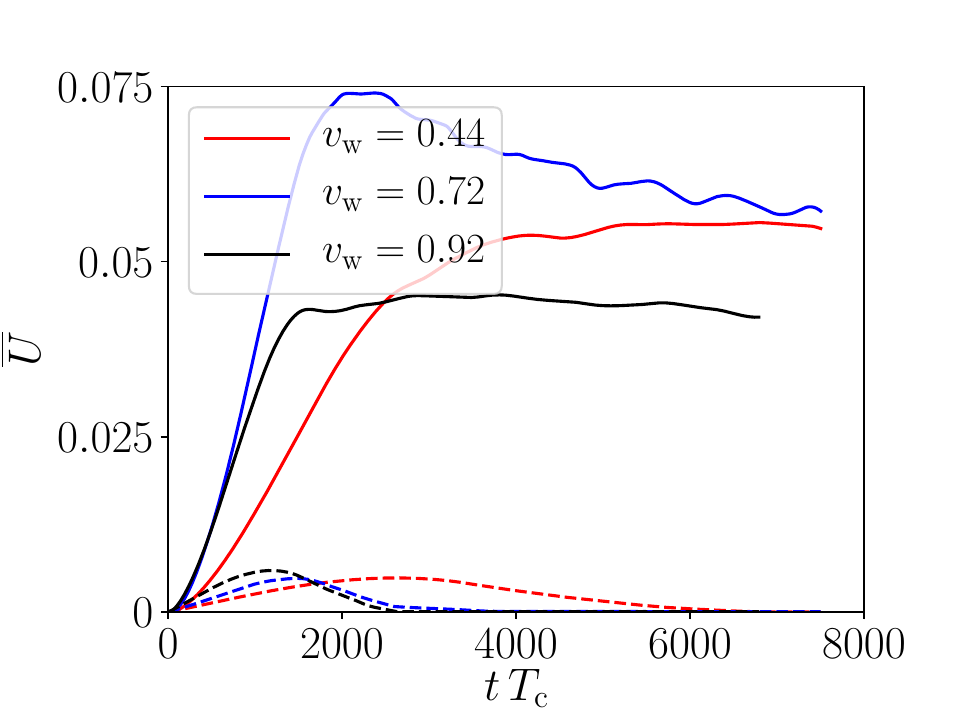}}
  \caption{
\label{fig:ubars}
RMS velocity and scalar gradient energy $\fluidV$ and $\fieldV$.
Solid lines denote the RMS fluid velocity, dashed lines the RMS scalar
field gradients. The simulations are separated into three plots
(a)-(c) according to phase transition strength and bubble radius.}
\end{figure*}

In Fig.~\ref{fig:ubars} we plot $\fluidV$ and $\fieldV$ against time,
showing the development, completion, and aftermath of the phase
transition.  We divide the transition into three phases
\cite{Hindmarsh:2015qta}: the expansion phase before any bubble
collisions take place, the collision phase, and the acoustic phase.
These can be traced in the Figures, using the fact that $\fieldV^2$ is
proportional to the surface area of the phase boundaries.  In the
expansion phase, $\fieldV$ grows linearly with time.  In the collision
phase, $\fieldV$ falls below the initial linear behaviour, peaks, and
drops to zero, marking the start of the acoustic phase. The RMS
velocity $\fluidV$ also grows with $\fieldV$ in the expansion and
collision phase, and levels off in the acoustic phase.  We take the
peak of $\fieldV$ to mark the collision time of the bubbles $\tColl$.

\begin{figure*}
  \subfigure[\ Weak, $\Rbc=1900/\Tc$]{\includegraphics[width=0.32\textwidth]{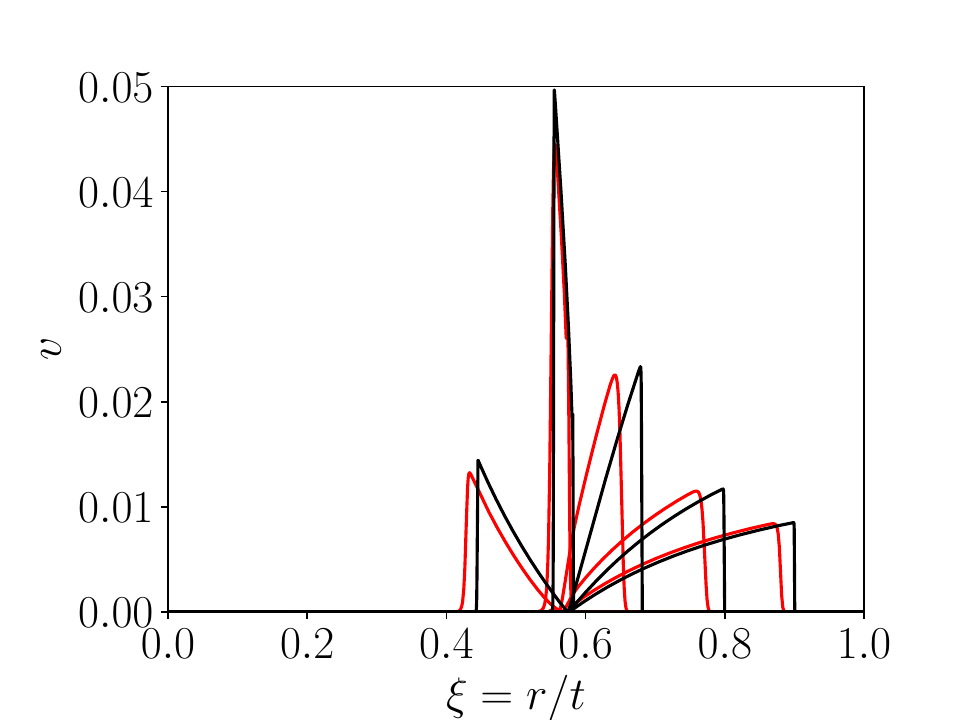}} \hfill
  \subfigure[\ Weak, $\Rbc=480/\Tc$]{\includegraphics[width=0.32\textwidth]{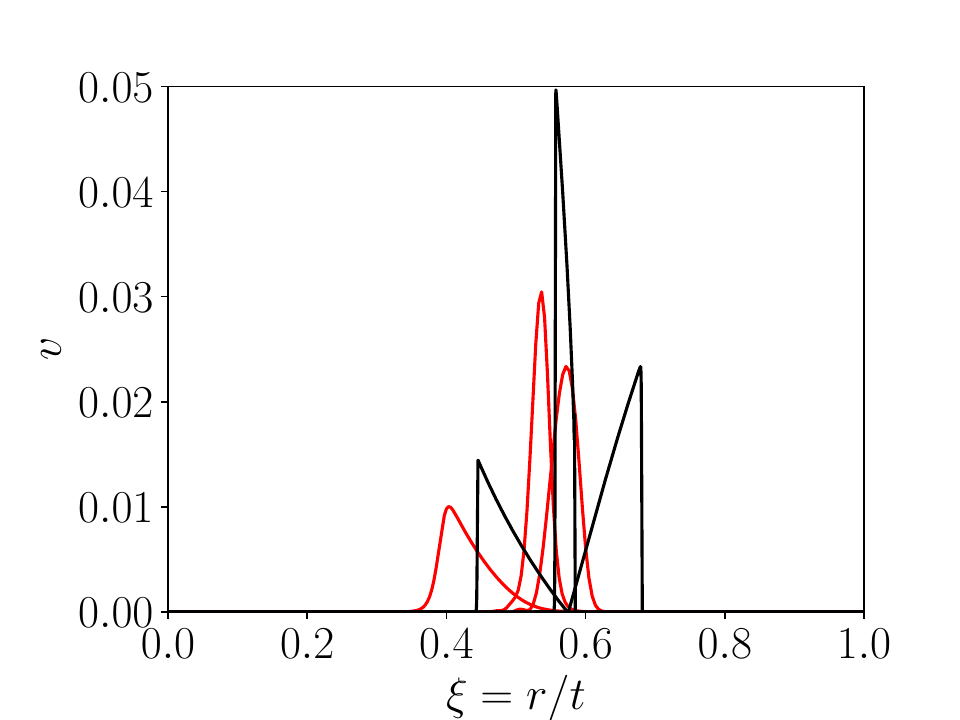}} \hfill
  \subfigure[\ Intermediate, $\Rbc=1900/\Tc$]{\includegraphics[width=0.32\textwidth]{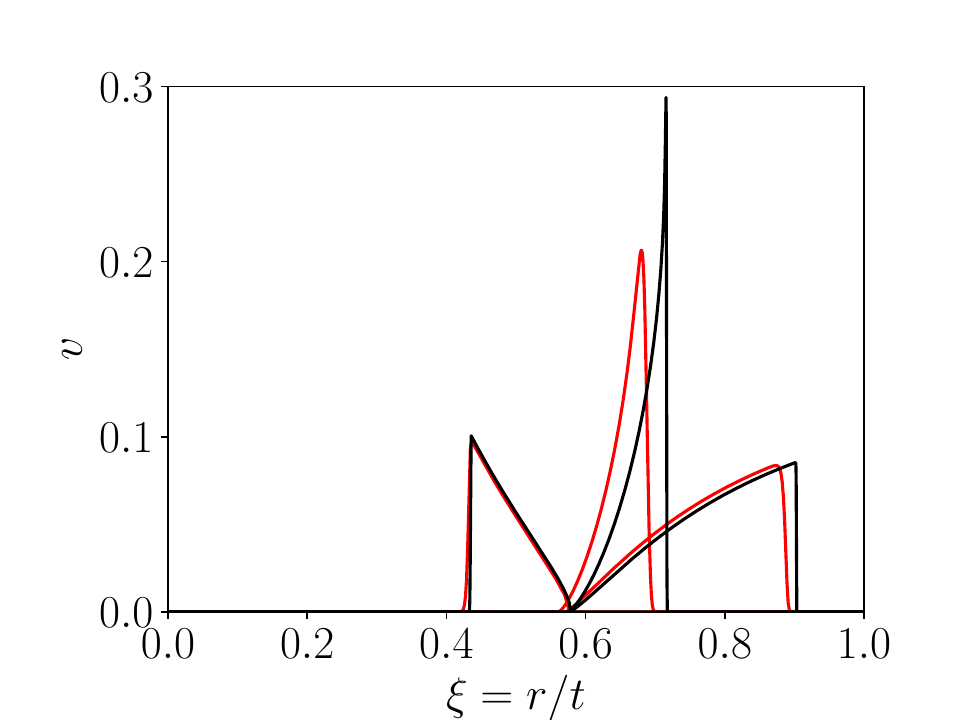}}
  \caption{
    \label{fig:fluidprofiles}
 Fluid radial velocity profiles as a function of scaled radius
 $\xi=r/t$. In red are curves taken at the peak of $\fieldV$ (see
 Fig.~\ref{fig:ubars}), at times $\tColl$ given in Table
 \ref{t:SimVelStats}.  In black are fluid velocities at late times, $t
 \gtrsim 10000/\Tc$.  Note that the wall speeds can be read off from
 the positions of the phase transition fronts.  Although the $\vw$
 quoted in the main text comes from simulations with a smaller lattice
 spacing ($dx \Tc = 0.2$), the discrepancy is small -- at most 3\% for
 the fastest detonations.}
\end{figure*}

The RMS fluid velocity is generally constant in the acoustic phase, as
noted in Ref.~\cite{Hindmarsh:2015qta}, although some reduction can be
seen in the transitions with bubbles expanding at near the
Chapman-Jouguet speed, where the fluid velocity profile is narrower
and peaks at higher values (see deflagrations with $\vw=0.56$ and
detonations with $\vw \simeq 0.7$ in Fig.~\ref{fig:fluidprofiles}).
It is possible that this reduction represents the beginning of
turbulent transport of energy to the lattice dissipation scale; the
time for shocks and turbulence to appear is in these cases $\tShock
\sim 10^5/\Tc$, only a factor of order 10 longer than the simulation
time.

Fig.~\ref{fig:fluidprofiles} also shows how the fluid profiles evolve
as the bubbles expand. The profile takes some time to settle to its
asymptotic self-similar form, and we plot both this, and the profile
at the peak collision time $\tColl$.  The difference is particularly
noticeable for small bubble separations where collisions happen much
earlier (see Fig.~\ref{fig:fluidprofiles}b); the fluid profiles are
smooth and do not have the characteristic sharp edges at the wall
position.  This affects the high wavenumber behaviour of the velocity
power spectra, as we shall see.

In Table \ref{t:SimVelStats} we list the maximum RMS fluid velocity
$\fluidVmax$, along with the transverse component $\fluidVmaxperp$.
We also give theoretical values for the mean square fluid velocity
estimated in two ways.  Firstly, $\fluidV^\mathrm{1D}$ is obtained by
integrating the numerical 1D fluid profiles out to $t=7000/\Tc$,
according to (see Refs.~\cite{Espinosa:2010hh,Hindmarsh:2015qta})
\begin{equation}
  \left({\fluidV^\mathrm{1D}}\right)^2 = \frac{3}{\vw^3} \int d\xi \, \xi^2 W^2 v^2,
\end{equation}
where $\xi=r/t$ is the scaled radius, $v(\xi)$ is the radial fluid
velocity and $W(\xi)$ the associated fluid gamma factor.

The second estimate is $\fluidV^\mathrm{Esp} =
\sqrt{\frac{3}{4}\kappa_\text{v}\al}$, where the function
$\kappa_\text{v}(\vw,\alpha)$ is given in the Appendix of Espinosa et
al. \cite{Espinosa:2010hh}, using $\vw$ extracted from 1D simulations
at $t = 7000/\Tc$, and $\al = \StrParB(\TN)$ from Table
\ref{t:SimParsPot}.  Note that $\kappa_\text{v}$ is defined from the
trace of the spatial part of the energy-momentum tensor as
\begin{equation}
\label{e:KapDef}
\kappa_\text{v} \StrParB = \frac{1}{\bar{\ep}\Vol } \int d^3 x \,
\fluidT^{ii}.
\end{equation}

\begin{table} 
\begin{tabular}{lccccccc}
    Type    &  {$\vw$} & {$\Rbc\Tc$} & $\tColl\Tc$ & $10^3 \fluidVmax$ & $10^3 \fluidVmaxperp$ &  $10^3 \fluidV^\mathrm{1D}$  &  $10^3 \fluidV^\mathrm{Esp}$   \\
\hline
Weak & 0.92 & 1918 & 1210 & 4.60 & 0.0833 & 5.31 & 5.32 \\
 & 0.80 & 1918 &  1380 & 5.75 & 0.0665 & 6.39 & 6.50 \\
 & 0.68 & 1918 & 1630 & 8.65 & 0.116 & 9.17 & 10.0 \\
 & 0.56 & 1889 & 1860 & 13.8 & 0.190 & 14.3 & 14.7 \\
 & 0.44 & 1918 & 2520 & 7.51 & 0.0775 & 7.70 & 7.76 \\
\cline{2-8}
 & 0.68 & 480 & 430 & 8.74 & 0.252 & 9.17 & 10.0 \\
 & 0.56 & 480 & 480 & 11.7 & 0.498 & 14.3 & 14.7 \\
 & 0.44 & 480 & 660 & 6.99 & 0.131 & 7.70 & 7.76 \\
\hline
Int. & 0.92 & 1918 & 1180 & 43.7 & 0.869 & 51.6 & 53.7 \\
 & 0.72 & 1889 & 1480 & 65.0 & 1.97 & 72.8 & 95.0 \\
 & 0.44 & 1918 & 2650 & 54.5 & 2.77 & 51.7 & 67.7 \\
\hline
\end{tabular}
\caption{\label{t:SimVelStats} Wall speed $\vw$, average bubble
  separation $\Rbc$, with peak bubble collision time $\tColl$, the
  maximum fluid RMS velocity $\fluidVmax$, the maximum contribution of
  transverse fluid motion $\fluidVmaxperp$, and two estimates for
  $\fluidV$ based on 1D fluid profiles and a fitting formula given in
  Ref.~\cite{Espinosa:2010hh}.  }
\end{table}

We see that the rotational component of the velocity field is very
small, consistent with the interpretation of the fluid flow as sound
waves, and with the linearity of the flow.  We also see that the
Espinosa et al.\ \cite{Espinosa:2010hh} fitting formula for the mean
square fluid velocity around a single bubble gives a good estimate of
the mean square velocity of the 3D flow.  Finally, the time at which
the scalar gradient energy peaks is approximately $\tColl \simeq 0.6
\Rbc/\vw$ in the simulations where $\Rbc$ is large.  This is
consistent with interpreting this peak as a bubble collision time.

\begin{figure*}
  \subfigure[\ Weak, $\vw = 0.92$]{\includegraphics[width=0.49\textwidth,clip=true]{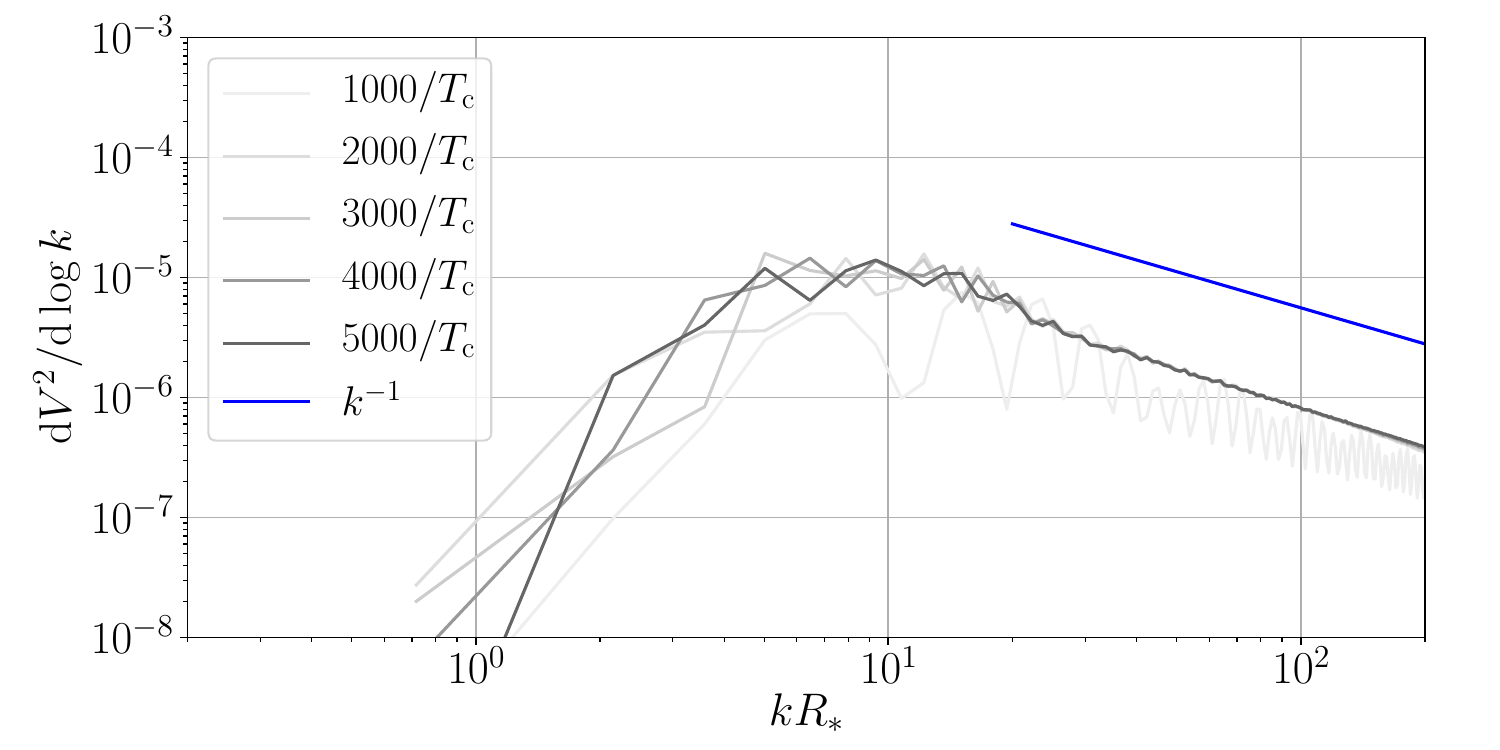}}
  \hfill	
  \subfigure[\ Intermediate, $\vw=0.92$]{\includegraphics[width=0.49\textwidth, clip=true]{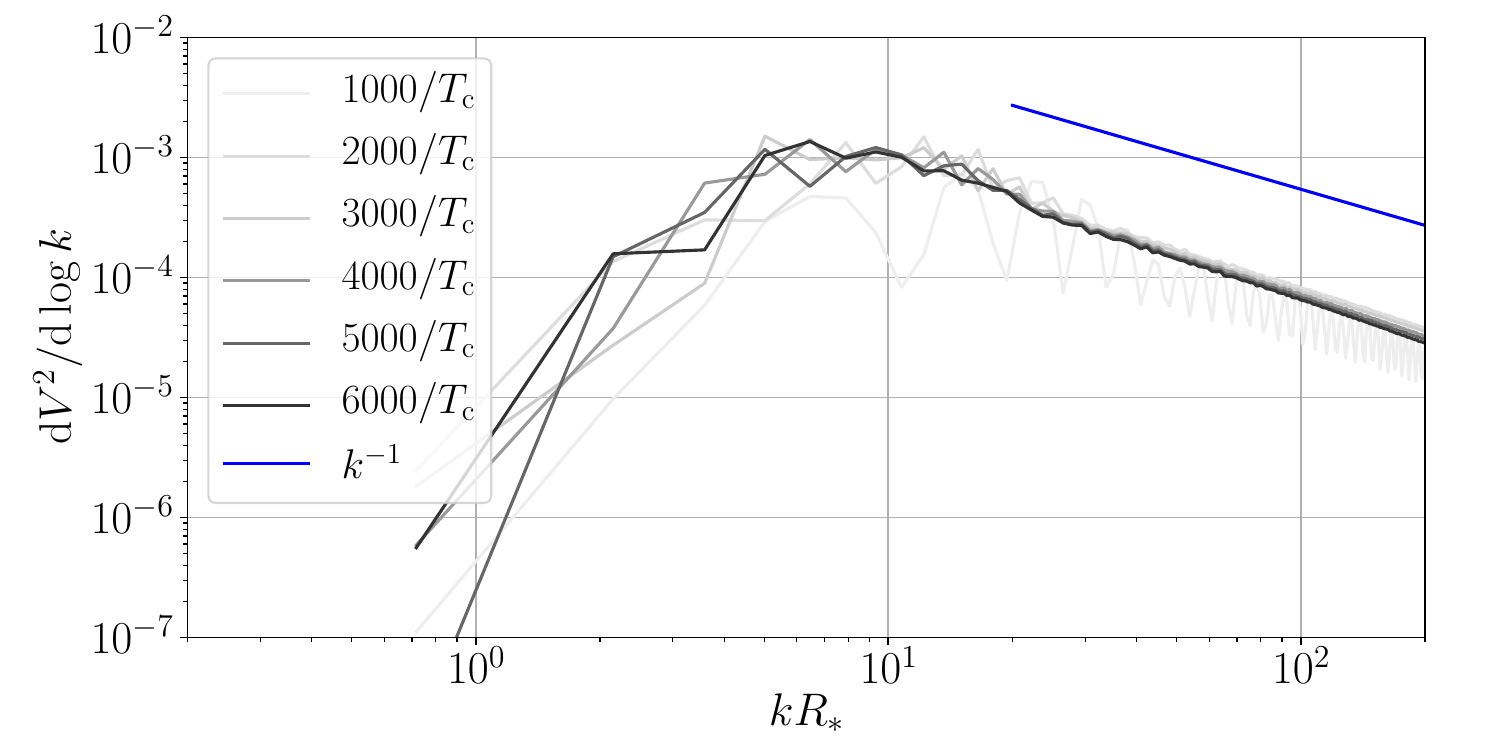}}
  \\
  \flushleft \subfigure[\ Weak, $\vw=0.8$]{\includegraphics[width=0.49\textwidth,clip=true]{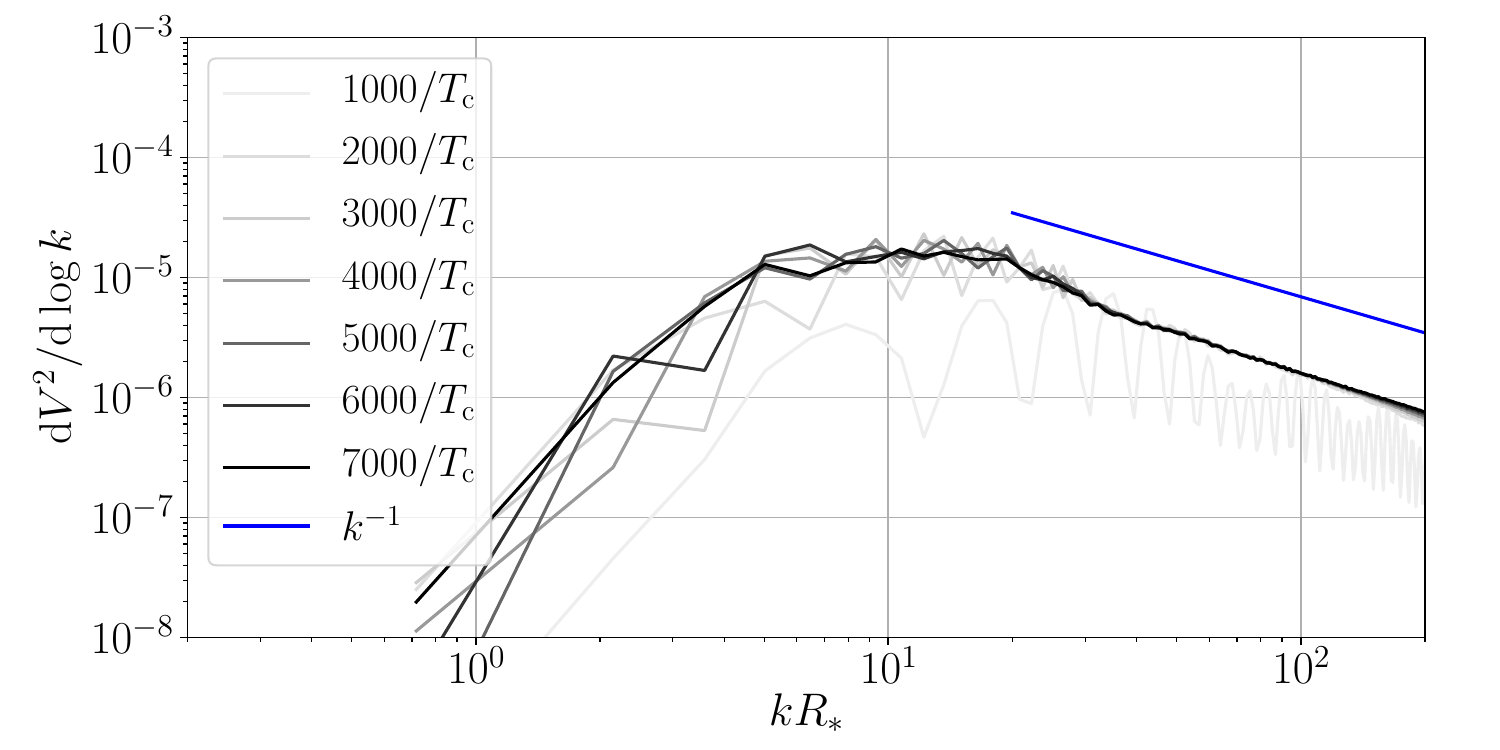}}
  \\
  \subfigure[\ Weak, $\vw=0.68$]{\includegraphics[width=0.49\textwidth,clip=true]{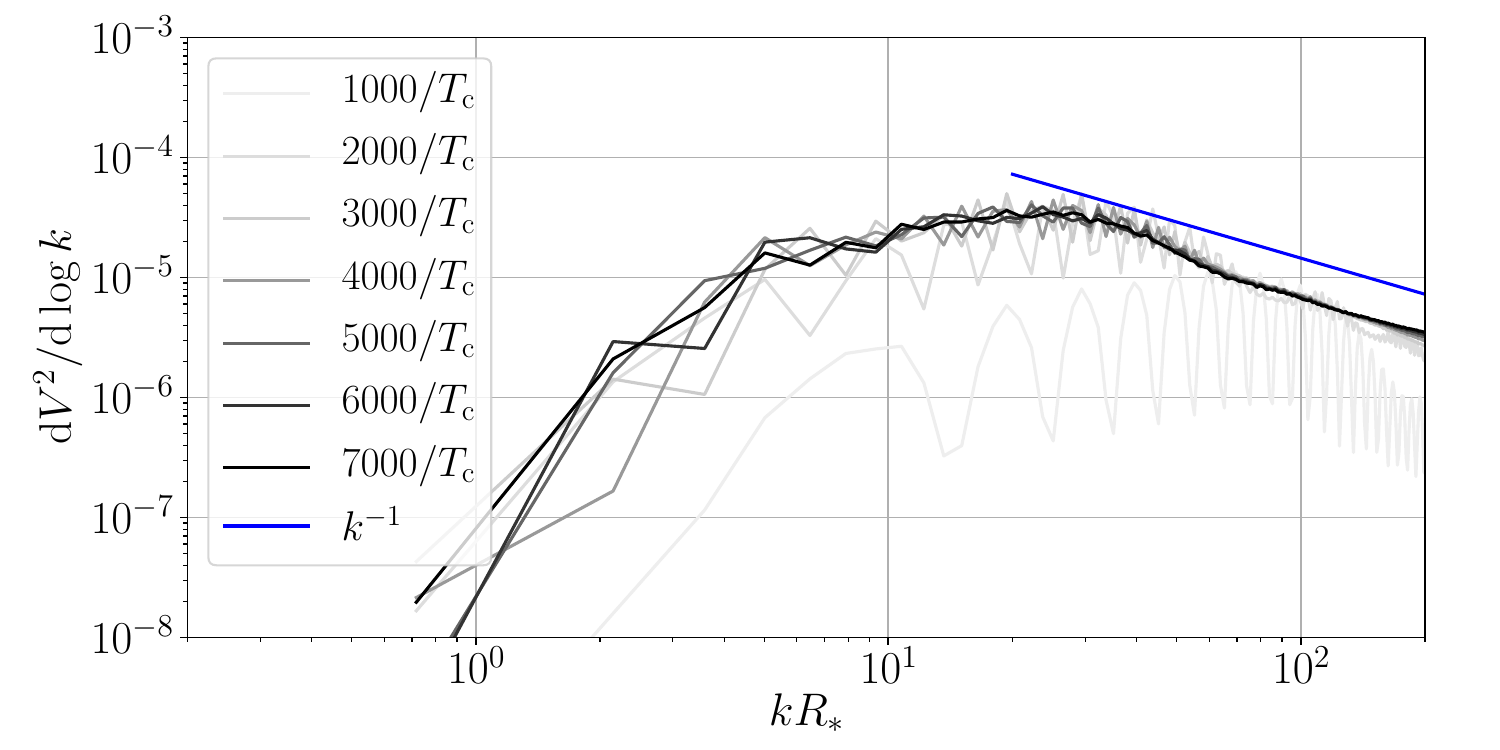}}
  \hfill
  \subfigure[\ Intermediate, $\vw=0.72$, $\Rbc=1889/\Tc$, $dx=1$]{\includegraphics[width=0.49\textwidth,clip=true]{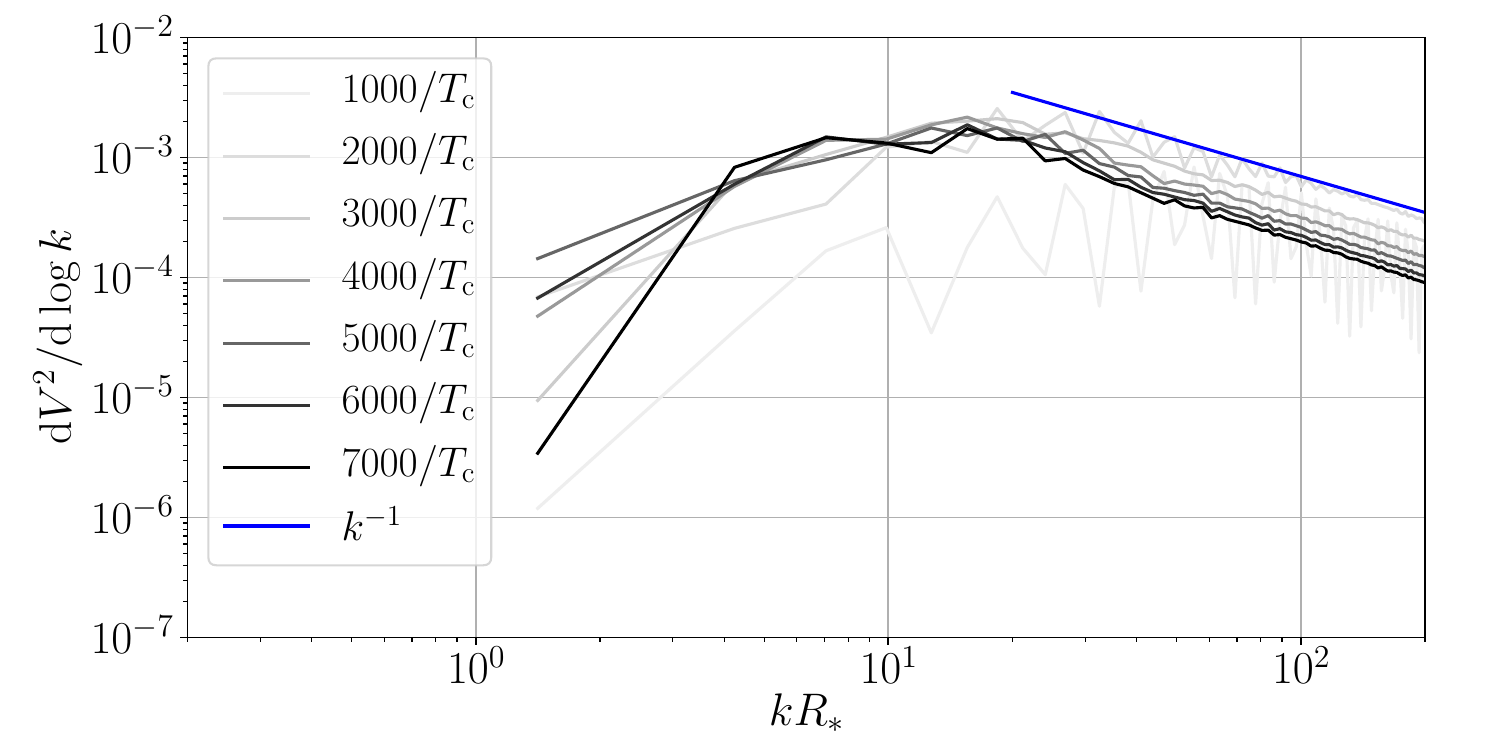}}

\caption{\label{f:VelPSDet} Velocity power spectra for detonations.
  Left are weak strength phase transitions, with $\vw = 0.92$, $0.80$
  and $0.68$.  Right are intermediate phase transitions, with $\vw =
  0.92$ and $\vw = 0.72$.  All have $N_b = 84$ bubbles (average
  separation $\Rc = 1918/\Tc$) and a lattice spacing $dx = 2/\Tc$,
  with the exception of the $\vw=0.72$ intermediate transition which
  has $N_b = 11$ ($\Rbc = 1889/\Tc$) and $dx = 1/\Tc$. Note there is
  no intermediate strength transition with $\vw=0.80$.  }
\end{figure*}

\begin{figure*}
\subfigure[\ Weak, $\vw=0.44$]{\includegraphics[width=0.49\textwidth, clip=true]{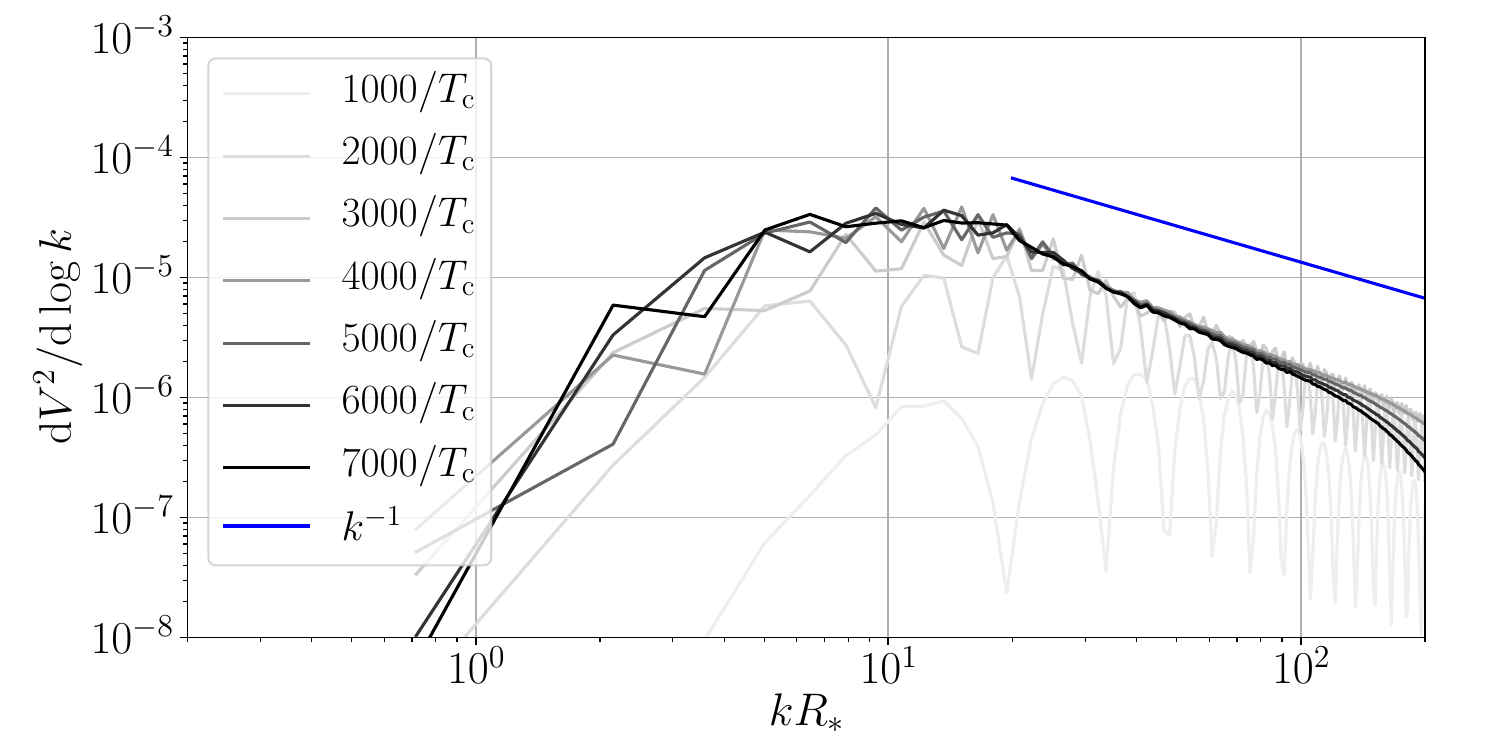}}
\hfill	
\subfigure[\ Intermediate, $\vw=0.44$]{\includegraphics[width=0.49\textwidth, clip=true]{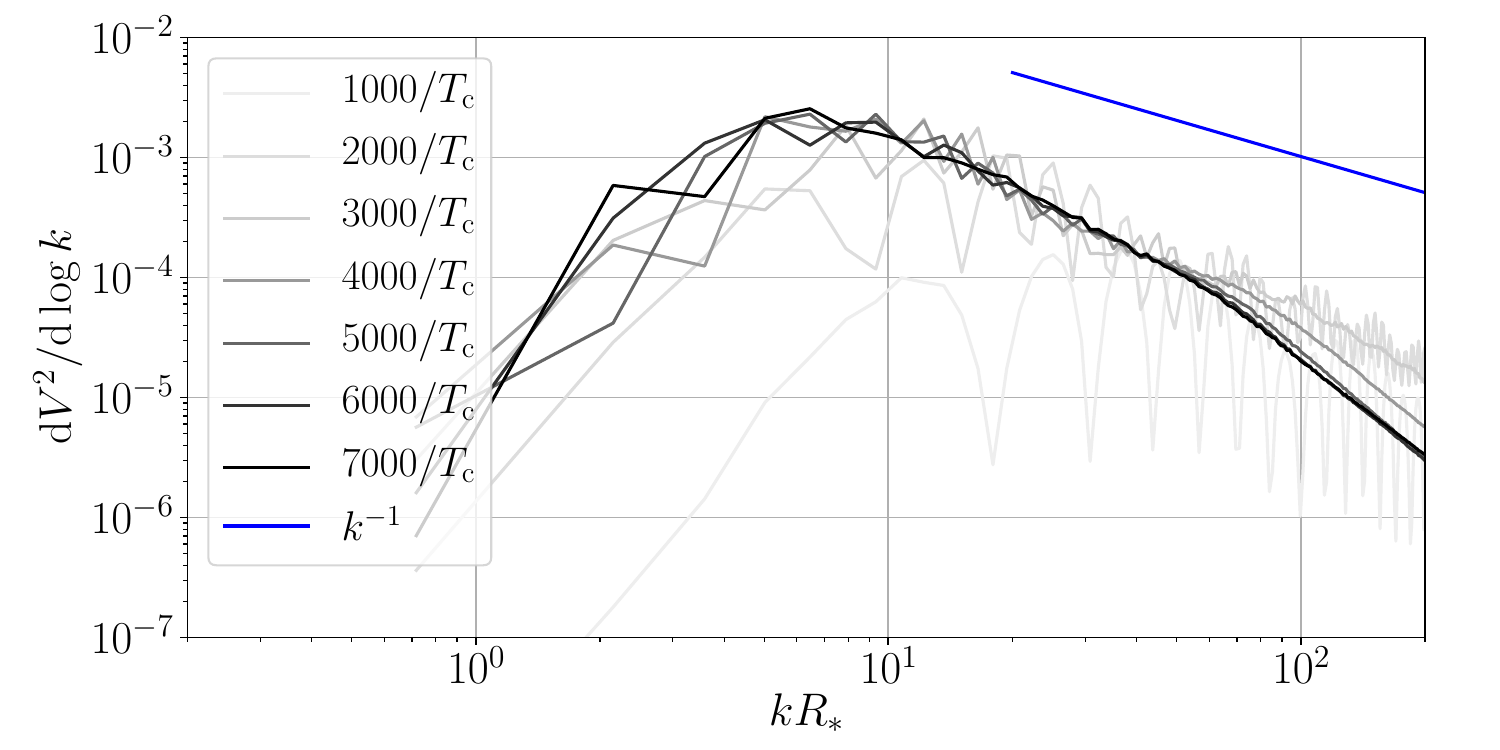}}\\

\caption{\label{f:VelPSDef}
Velocity power spectra for deflagrations with $\vw=0.44$. 
Left is a weak phase transition, right is an intermediate transition. 
Both have $N_b = 84$ bubbles (mean separation $\Rc = 1918/\Tc$).
}

\end{figure*}

\begin{figure}
\flushleft
\subfigure[\ $\Nb=11$, $dx=1$]{\includegraphics[width=0.49\textwidth,clip=true]{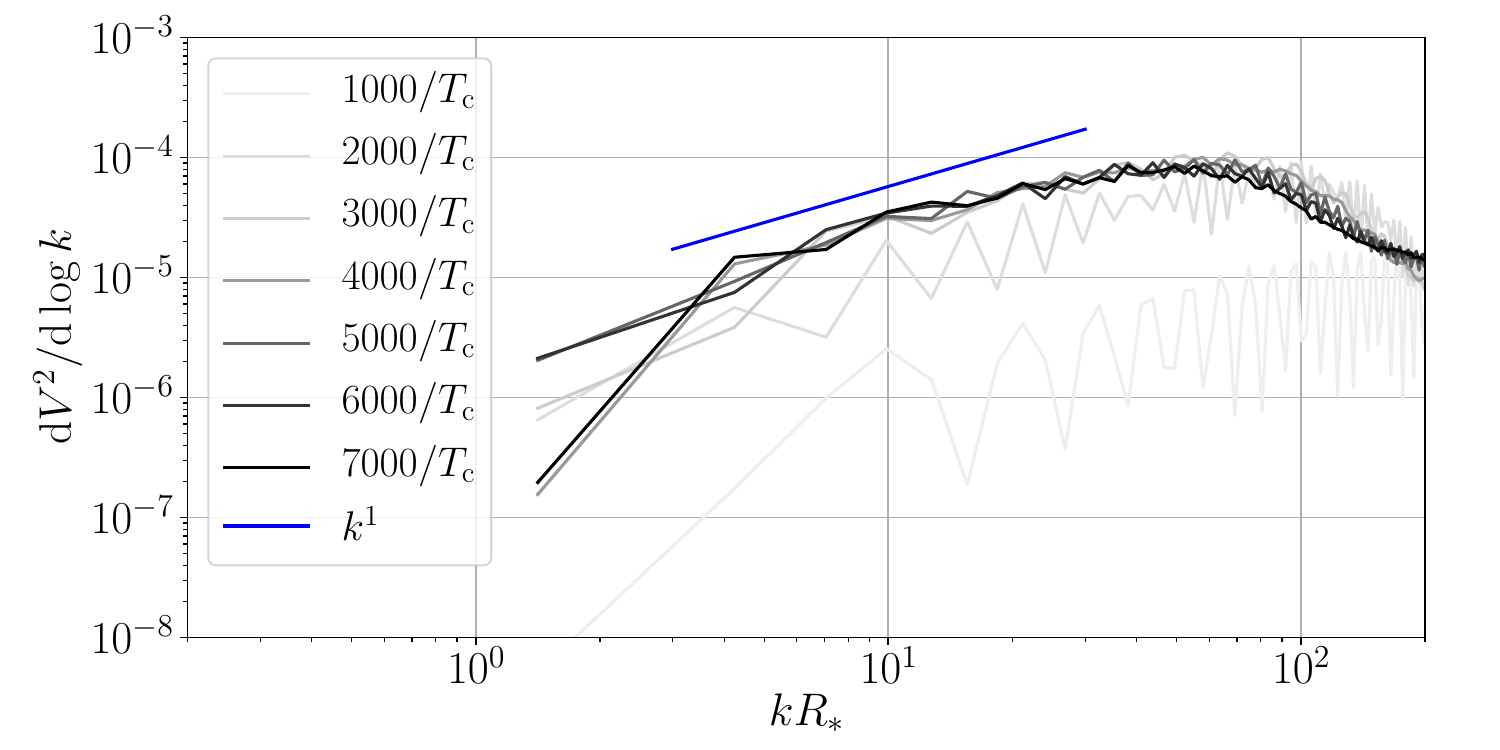}}
\hfill	
\\
\flushleft
\subfigure[\ $\Nb=84$]{\includegraphics[width=0.49\textwidth,clip=true]{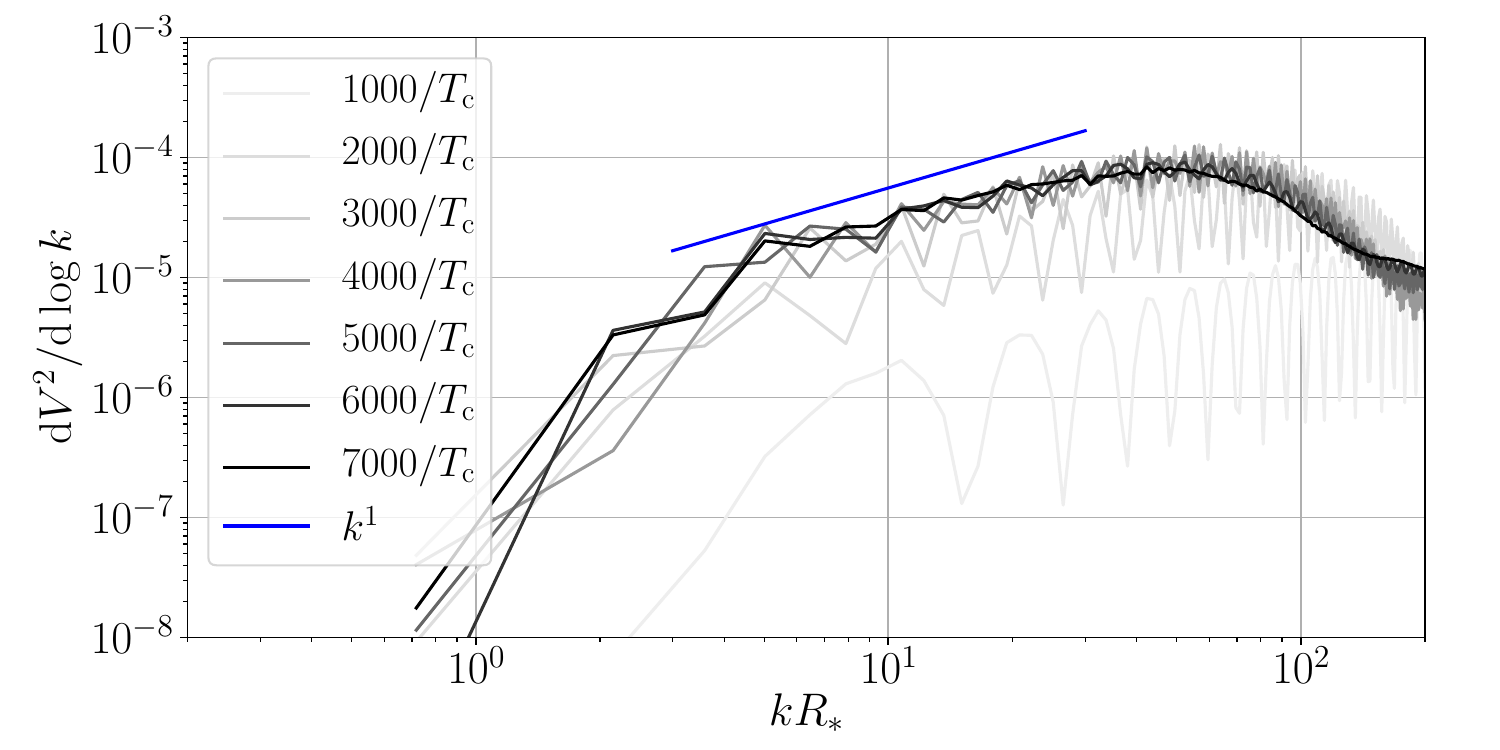}}
\hfill	
\\
\flushleft
\subfigure[\ $\Nb=5376$]{\includegraphics[width=0.49\textwidth,clip=true]{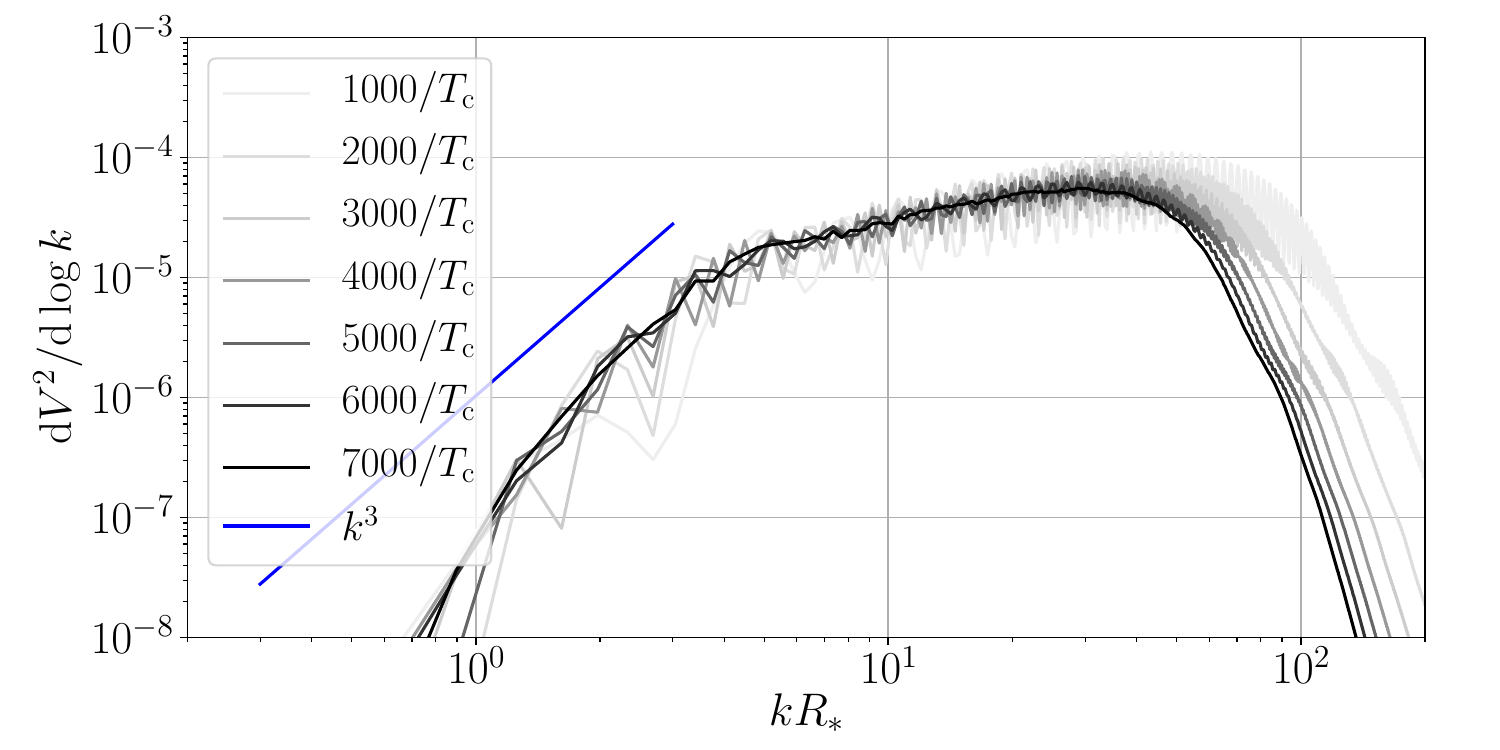}}

\caption{\label{f:VelPSDefJou}
Velocity power spectra for near-Jouguet deflagrations.
All are weak transitions with $\vw=0.56$, and $N_b = 11, 84$ and $5376$ ($\Rc = 1889/\Tc$, $1918/\Tc$, and $480/\Tc$).
}

\end{figure}

In Figs.~\ref{f:VelPSDet}, \ref{f:VelPSDef} and \ref{f:VelPSDefJou} we
plot velocity power spectra (\ref{e:VPowSpe}) for a set of
simulations.  We divide the data according to the bubble wall speed
$\vw$, separating out the special case of deflagrations moving at
close to sound speed.

In Fig.~\ref{f:VelPSDet} we show detonations with wall speeds $\vw =
0.92$, $0.80$ and $\vw \simeq 0.7$ for transitions of weak and
intermediate strength.  The intermediate strength transition at $\vw =
0.72$ is run at a higher resolution ($dx=1/\Tc$) to resolve the higher
velocity gradients, so that the mean bubble separation ($1889/\Tc$) is
close to the box size of the simulation ($2100/\Tc$), hence there is
less dynamic range on the long-wavelength side of the peak in the
power spectrum.

The general form is a broken power law, with a domed peak at $k\Rbc =
\textrm{O}(10)$.  The shape is similar between the weak and
intermediate cases at the same velocity; the intermediate strength
transitions have higher amplitude, as more energy is transferred into
kinetic energy.  This is in accord with the sound shell model
\cite{Hindmarsh:2016lnk}, where the velocity power spectra are of a
universal shape for a given wall velocity and non-relativistic fluid
flows with negligible shocks.

The width of the dome is larger for the detonations with wall speeds
closer to the speed of sound, with the peak displaced to the right
relative to the fast detonation.  This is also consistent with the
sound shell model prediction that the peak position in the power
spectrum is determined by the inverse width of the sound shell.

The power-law to the right of the dome is close to the $k^{-1}$
predicted by the sound shell model, particularly in the detonations,
where there is a clear separation between the peak scale and the wall
width scale.  The long-wavelength power law index is not so clear, as
there are few bins and there are fewer $k$ vectors in each bin, but is
consistent with the predicted $k^{3}$.

In Fig.~\ref{f:VelPSDef} we show velocity power spectra from
deflagrations with wall speed $\vw = 0.44$.  The power law to the
right of the dome appears steeper than the $k^{-3}$ prediction, and
there is a knee in the power spectrum at higher $k$ neither of which
is in accord with the sound shell model.  These features need further
investigation.

We recall that in the sound shell model, there are two scales in the
velocity field: the mean bubble separation $\Rbc$, and the sound shell
width $\Delta\Rbc = \Rbc \De \vw/\vw$.  If the scales are well
separated (i.e.\ if the bubble wall is moving close to the speed of
sound), the long-wavelength $k^3$ power law is predicted to turn into
a $k^{1}$ power law at $k\Rbc = \textrm{O}(1)$, and finally to a
$k^{-1}$ power law at $k\De\Rbc = \textrm{O}(1)$.

The clearest scale separation should be found in transitions where the
wall speed is closest to the sound speed.  In Fig.~\ref{f:VelPSDefJou}
we show the power spectra from a deflagration with speed $\vw = 0.56$,
very close to $\cs = 1/\sqrt{3} \simeq 0.577$.

We do not have the dynamic range to resolve all three wavelength
ranges simultaneously, but by altering the number of bubbles we can
try to resolve two ranges at a time.  At the top, for the largest
$\Rbc$, we see that the dome around the peak has broadened into a
slowly rising plateau, consistent with a $k^{1}$ behaviour.  At higher
wavenumber the plateau drops off, although we do not have enough range
to confirm a $k^{-1}$ behaviour.

For larger numbers of bubbles (centre, bottom in
Fig.~\ref{f:VelPSDefJou}) the long-distance behaviour emerges. With
$\Nb = 5376$ ($\Rbc = 480/\Tc$) one can clear see a steep power law,
even steeper than $k^{3}$, in the range $1 \lesssim k\Rbc \lesssim 3$.
A possible reason for the discrepancy with the generic sound shell
model prediction at low $k\Rbc$ is that the coefficient of the $k^3$
term is proportional to $\De \vw^2 \simeq 4\times10^{-4}$
\cite{Hindmarsh:2016lnk}, and so the next order term in a series
expansion in $k\Rbc$ can dominate for $k\Rbc = \mathrm{O}(1)$.  The
short-distance behaviour at these large bubble numbers is not
reliable, as there is insufficient distinction between the sound shell
width $\De\Rbc \simeq 15/\Tc$ and the bubble wall width $\ell \simeq
3/\Tc$. Indeed, in Fig.~\ref{fig:fluidprofiles} one can see that the
fluid velocity profiles are far from their asymptotic forms.

\section{Results: gravitational waves}
\label{sec:gws}

\begin{table} 
\begin{tabular}{lcccccc}
Type & $\vw$ & $\Rbc\Tc$ & $\IntSca_\text{f}^{\text{end}}\Tc $ & $\IntSca_\text{gw}^\text{end}\Tc$ & $10^2\OmGWscaled^{\IntSca_\text{f}}$ & $10^2\OmGWscaled^{\Rbc}$  \\
\hline
Weak & 0.92 & 1918 & 1490 & 1620 & 1.5 & 1.2  \\ 
 & 0.80 & 1918 & 1290 & 1600 & 2.2 & 1.4  \\ 
 & 0.68 & 1918 & 888 & 1410 & 1.5 & 0.62  \\ 
 & 0.56 & 1889 & 530 & 865 & 1.4 & 0.32  \\ 
 & 0.44 & 1918 & 1450 & 1750 & 1.4 & 1.1  \\ 
\cline{2-7}
 & 0.68 &  480 & 268 & 323 & 1.5 & 0.88  \\ 
 & 0.56 &  480 & 233 & 416 & 1.6 & 0.86  \\ 
 & 0.44 &  480 & 416 & 493 & 1.5 & 1.3  \\ 
\hline
Int. & 0.92 & 1918 & 1530 & 1780 & 2.6 & 2.0  \\ 
 & 0.72 & 1889 & 1100 & 1180 & 3.3 & 1.8  \\ 
 & 0.44 & 1918 & 1980 & 2090 & 1.6 & 1.7  \\ 
\hline
\end{tabular}
\caption{\label{t:SimGWs} Wall speed $\vw$ and mean bubble separation
  $\Rbc$, with the resulting integral scale of the fluid flow
  $\IntSca_\text{f}^{\text{end}}$, the integral scale of the
  gravitational wave power $\IntSca_\text{gw}^\text{end}$, and the
  dimensionless gravitational wave amplitude parameters 
  $\OmGWscaled^{\IntSca_\text{f}}$ and
$\OmGWscaled^{\Rbc}$. 
}
\end{table}

In Table \ref{t:SimGWs} we show global quantities computed from the
gravitational wave power spectrum
\begin{equation}
\PspecGW(z) = \frac{z^3}{2\pi^2}\SpecDenGW(z),
\end{equation}
where $z = k\Rbc$.
These are the integral scale of the gravitational waves 
\begin{equation}
\label{e:IntScaGW}
\IntSca_\text{gw} =  \frac{1}{\OmGW}\int \frac{dk}{k} \frac{1}{k} \frac{d \OmGW}{ d \ln(k)},
\end{equation}
calculated at the end of the simulations, and the dimensionless
gravitational wave amplitude parameter defined in (\ref{e:OmTilDef}),
with the fluid length scale $\fluidL$ is taken to be either the
integral scale of the velocity field $\IntSca_\text{f}$ (defined
analogously to Eq.~(\ref{e:IntScaGW})) or the mean bubble separation
$\Rbc$.  With either of these length scales, $\OmGWscaled$ is
approximately constant and of order $10^{-2}$ for weak
transitions\footnote{Note that there is a numerical error in the
  computation of $\OmGWscaled$ in \cite{Hindmarsh:2015qta} which
  resulted in a value which was a factor $(2\pi)^3/32\pi$ too
  high.}. The integral scales of both the fluid and the gravitational
waves are significantly smaller for the near-Jouguet transitions,
showing the influence of the sound shell thickness.

\begin{figure*}

\subfigure[\ Weak, $\vw=0.92$]{\includegraphics[width=0.49\textwidth,clip]{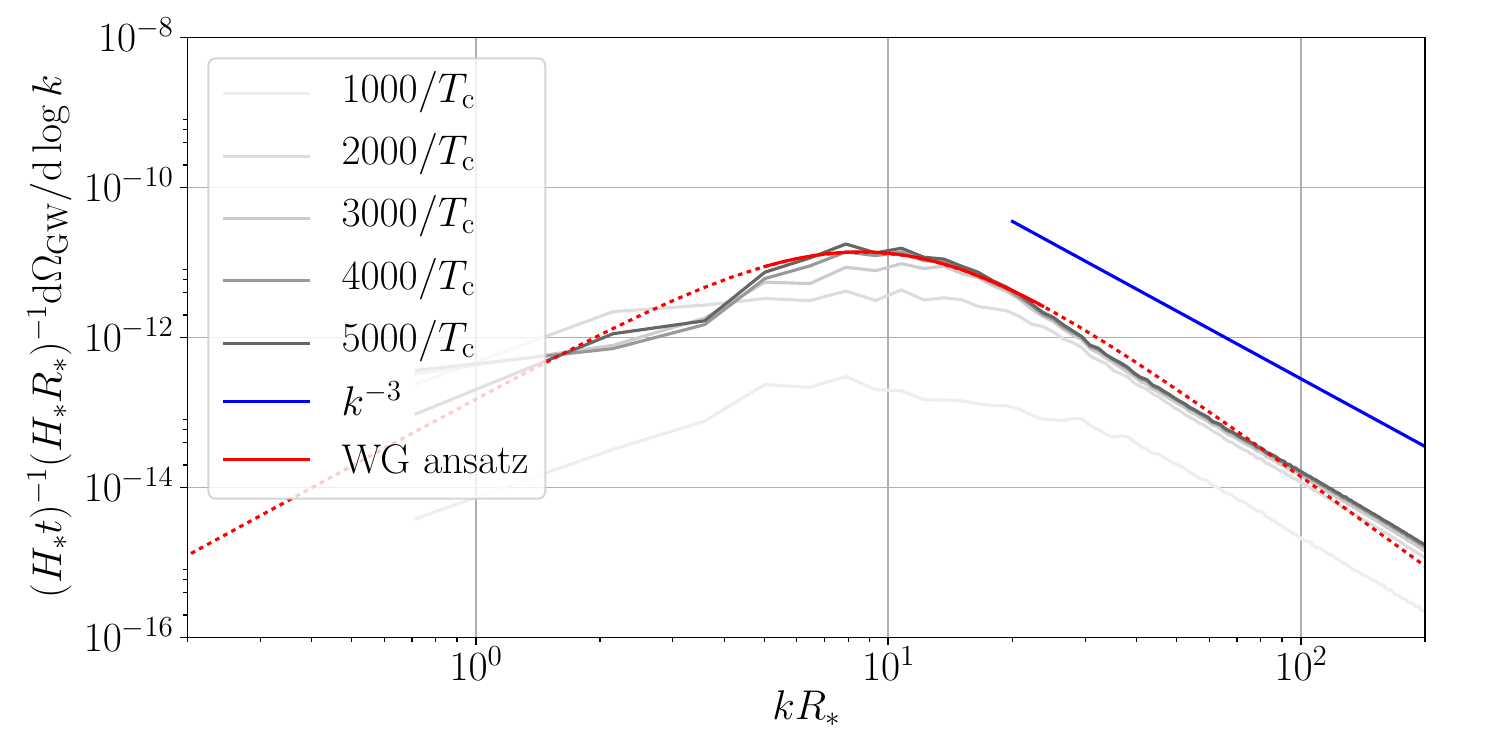}}
\hfill	
\subfigure[\ Intermediate, $\vw=0.92$]{\includegraphics[width=0.49\textwidth,clip]{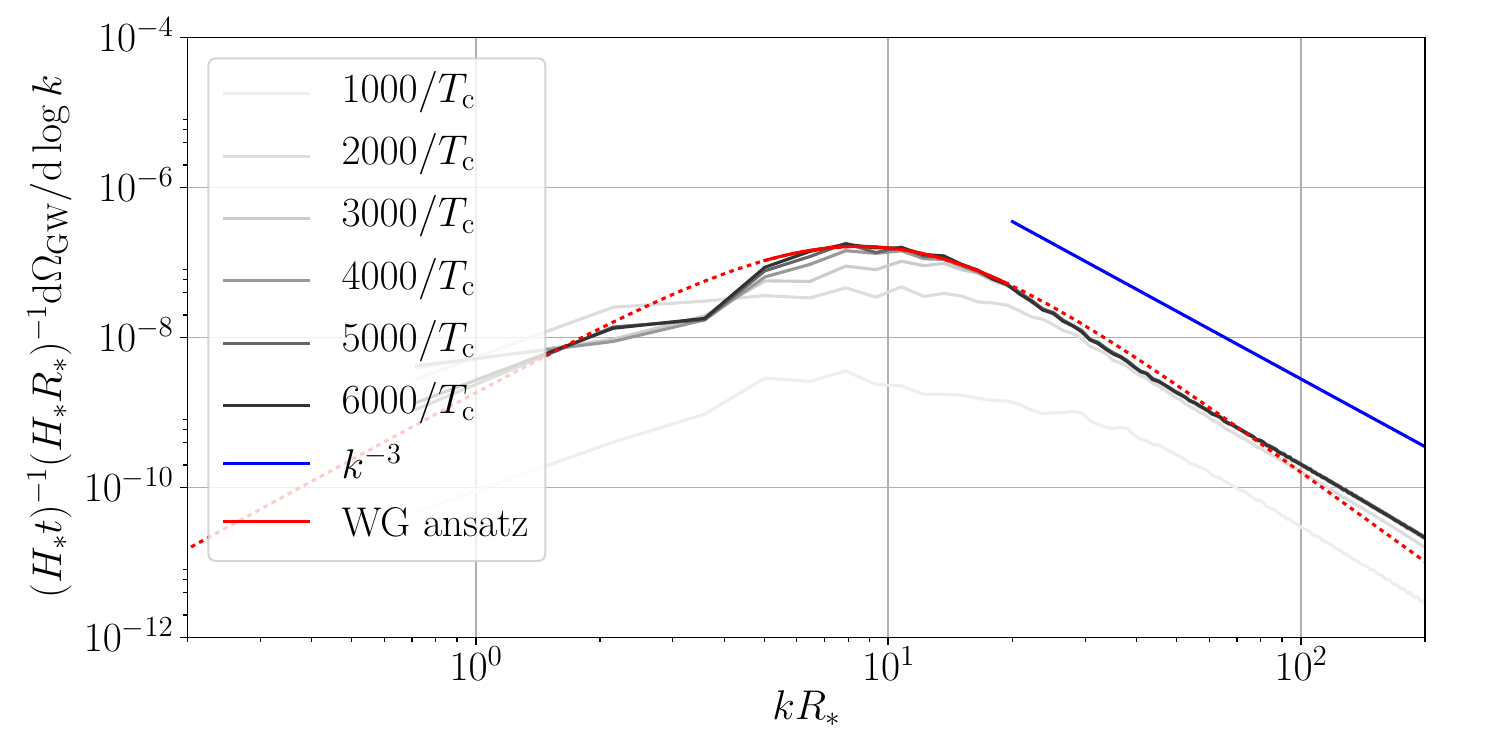}}\\
\flushleft
\subfigure[\ Weak, $\vw=0.8$]{\includegraphics[width=0.49\textwidth,clip]{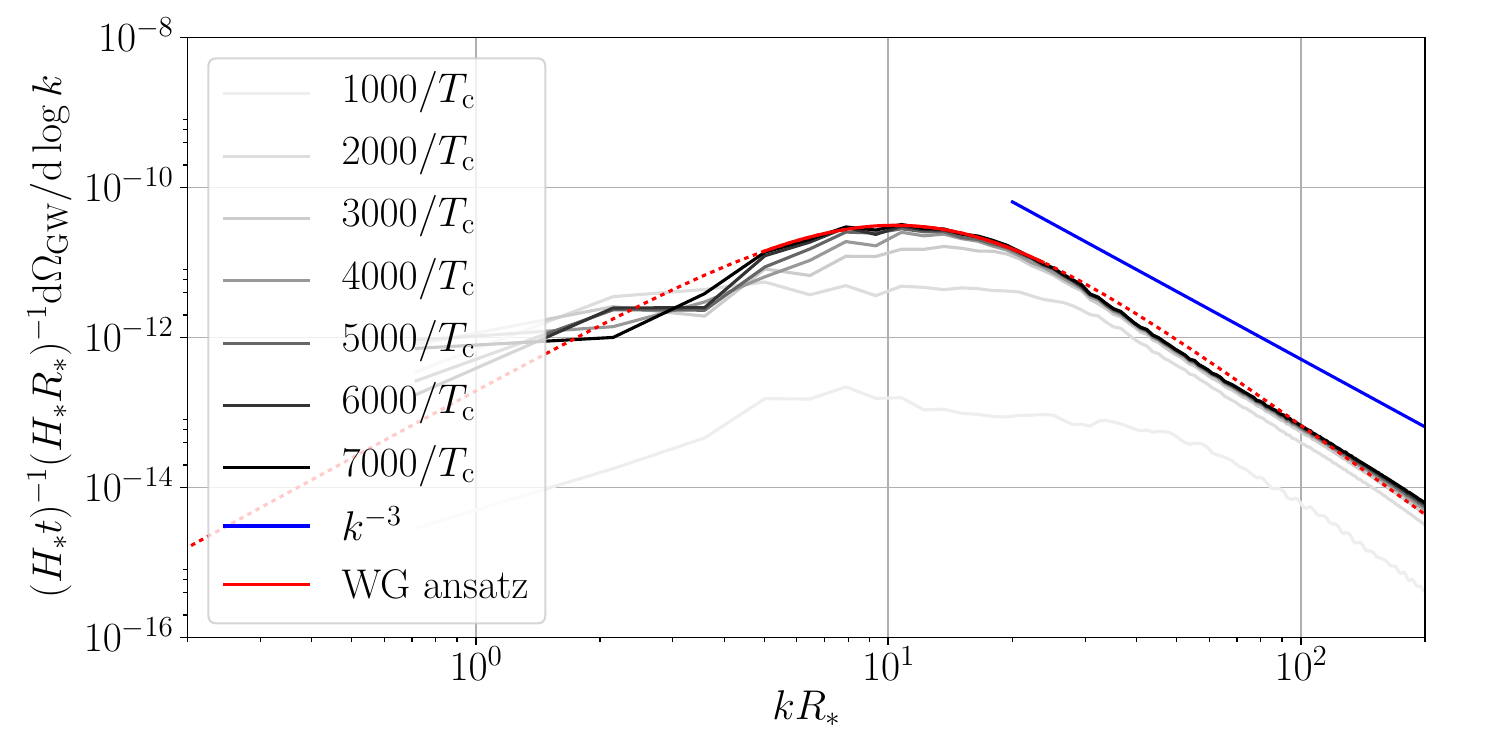}}
\\
\subfigure[\ Weak, $\vw=0.68$]{\includegraphics[width=0.49\textwidth,clip]{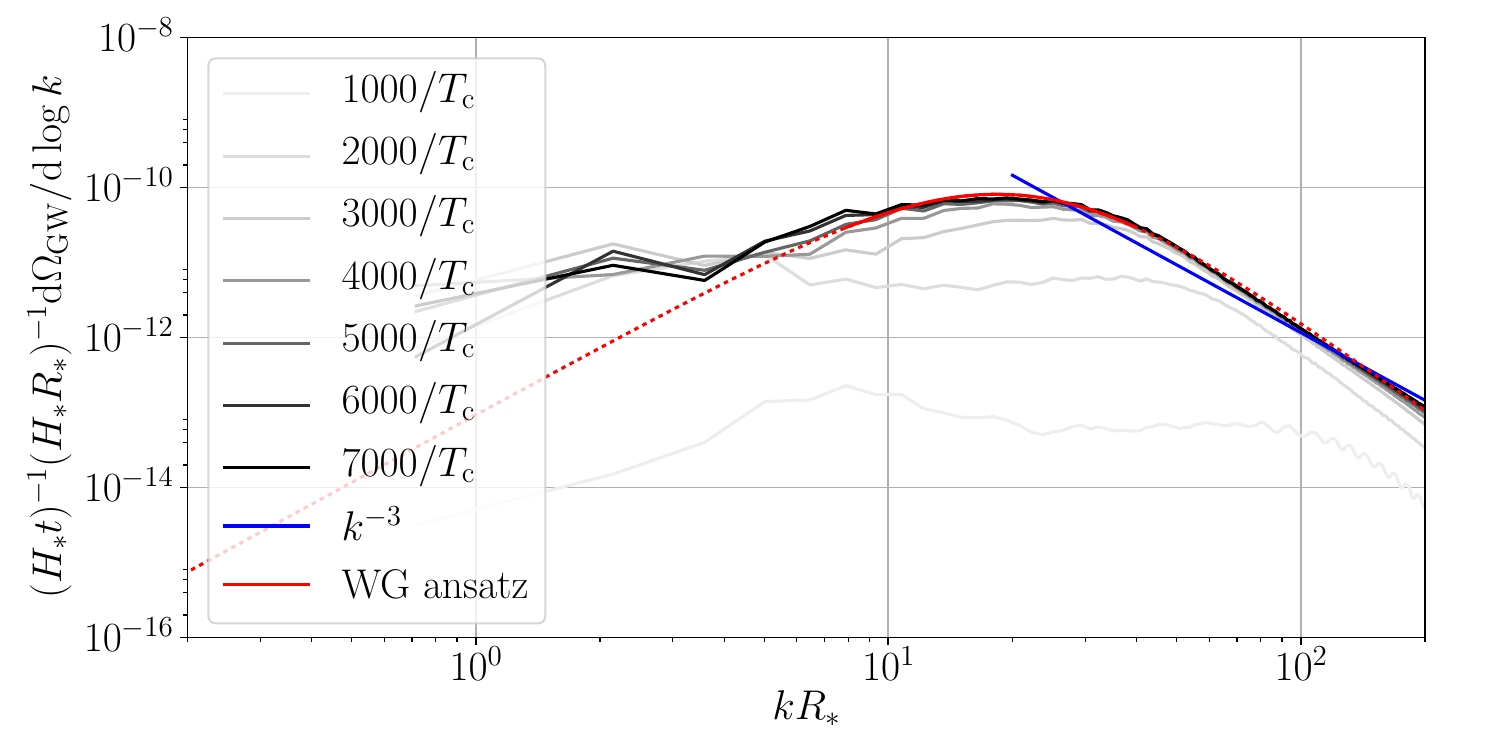}}
\hfill
\subfigure[\ Intermediate, $\vw=0.72$, $\Rbc = 1889/\Tc$, $dx = 1$]{\includegraphics[width=0.49\textwidth,clip]{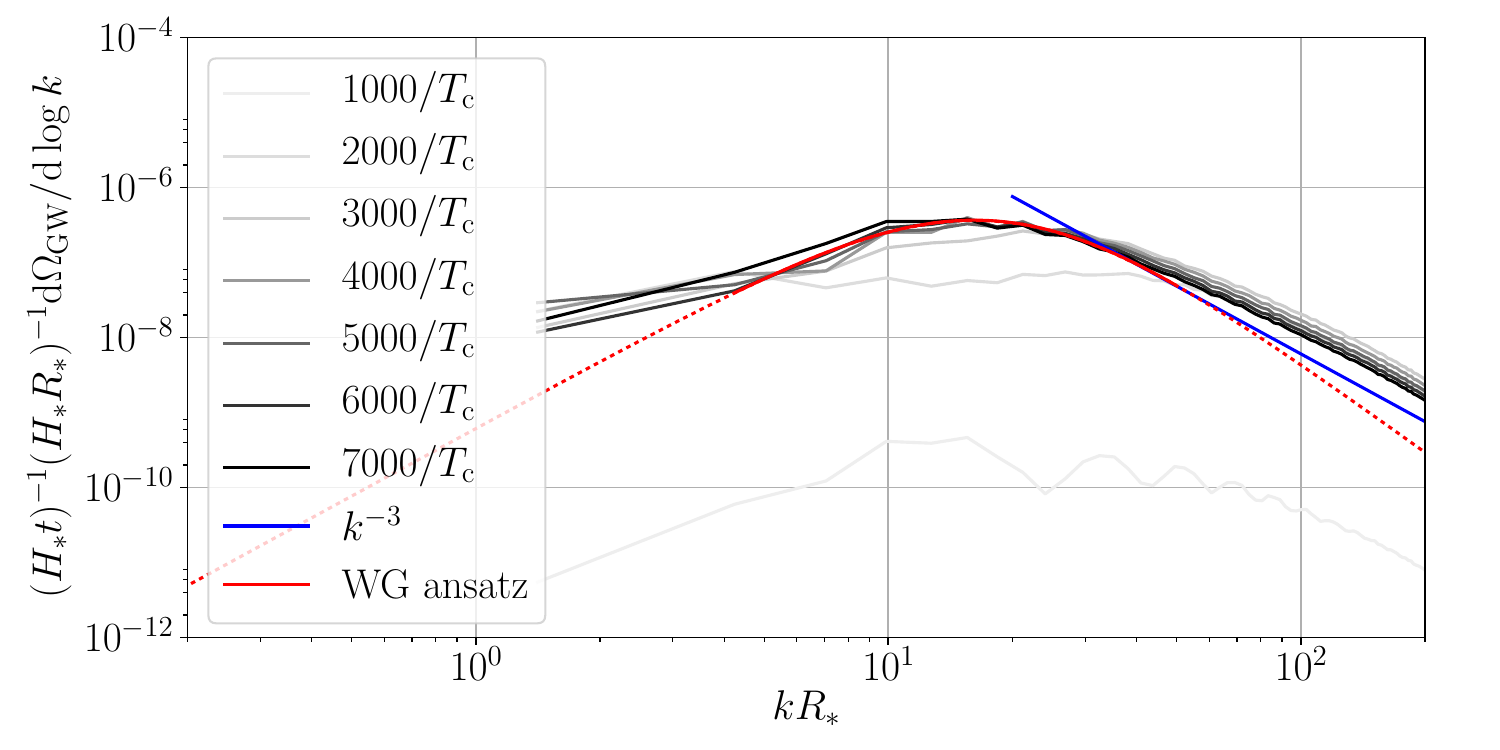}}

\caption{\label{f:GWPSDet} Power spectra of fractional energy density
  in gravitational waves for detonations, divided by the ratio of the
  mean bubble separation $\Rbc$ to the Hubble length at the transition
  $\HN$, and the ratio of the time to the Hubble time.  The wave
  number is scaled by the mean bubble separation.  Left are weak phase
  transitions, showing detonations with $\vw = 0.92$, $0.80$ and
  $0.68$ (top to bottom).  Right are intermediate phase transitions,
  with wall speeds $\vw=0.92$ and $\vw = 0.72$.  All have $N_b = 84$
  bubbles, giving a mean bubble separation of $\Rbc = 1918/\Tc$, and a
  lattice spacing $dx = 2/\Tc$, with the exception of the $\vw=0.72$
  intermediate transition which has $N_b = 11$ ($\Rc = 1889/\Tc$) and
  $dx = 1/\Tc$.  }

\end{figure*}

\begin{figure*}
\caption{\label{f:GWPSDef} Power spectra of fractional energy density
  in gravitational waves for deflagrations with $\vw = 0.44$, divided
  by the ratio of the mean bubble separation $\Rbc$ to the Hubble
  length at the transition $\HN$, and the ratio of the time to the
  Hubble time.  The wave number is scaled by the mean bubble
  separation.  Left is a weak phase transition, right is an
  intermediate transition.  Both have $\vw = 0.44$ and $N_b = 84$
  ($\Rbc = 1918/\Tc$).  } \subfigure[\ Weak,
  $\vw=0.44$]{\includegraphics[width=0.49\textwidth,
    clip=true]{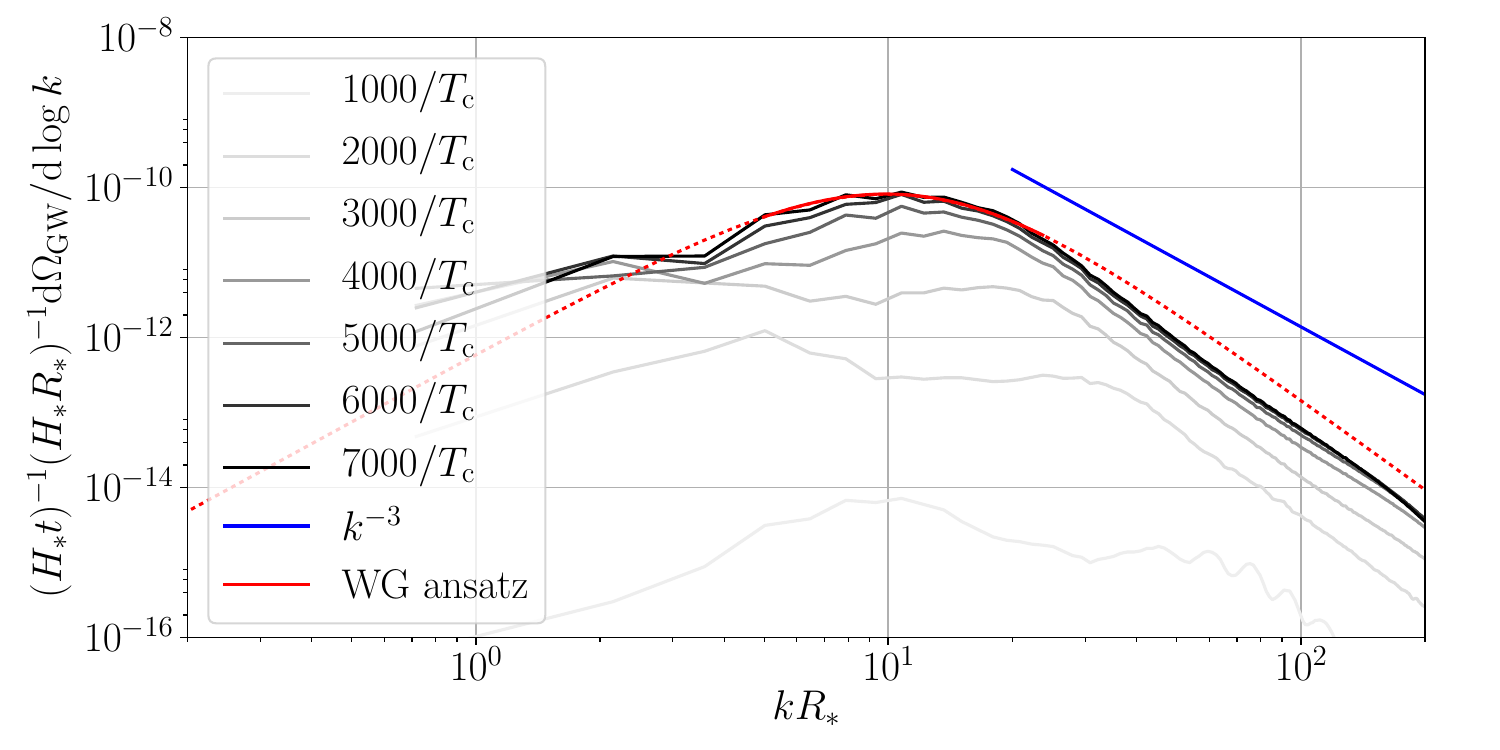}}
\hfill \subfigure[\ Intermediate,
  $\vw=0.44$]{\includegraphics[width=0.49\textwidth,clip=true]{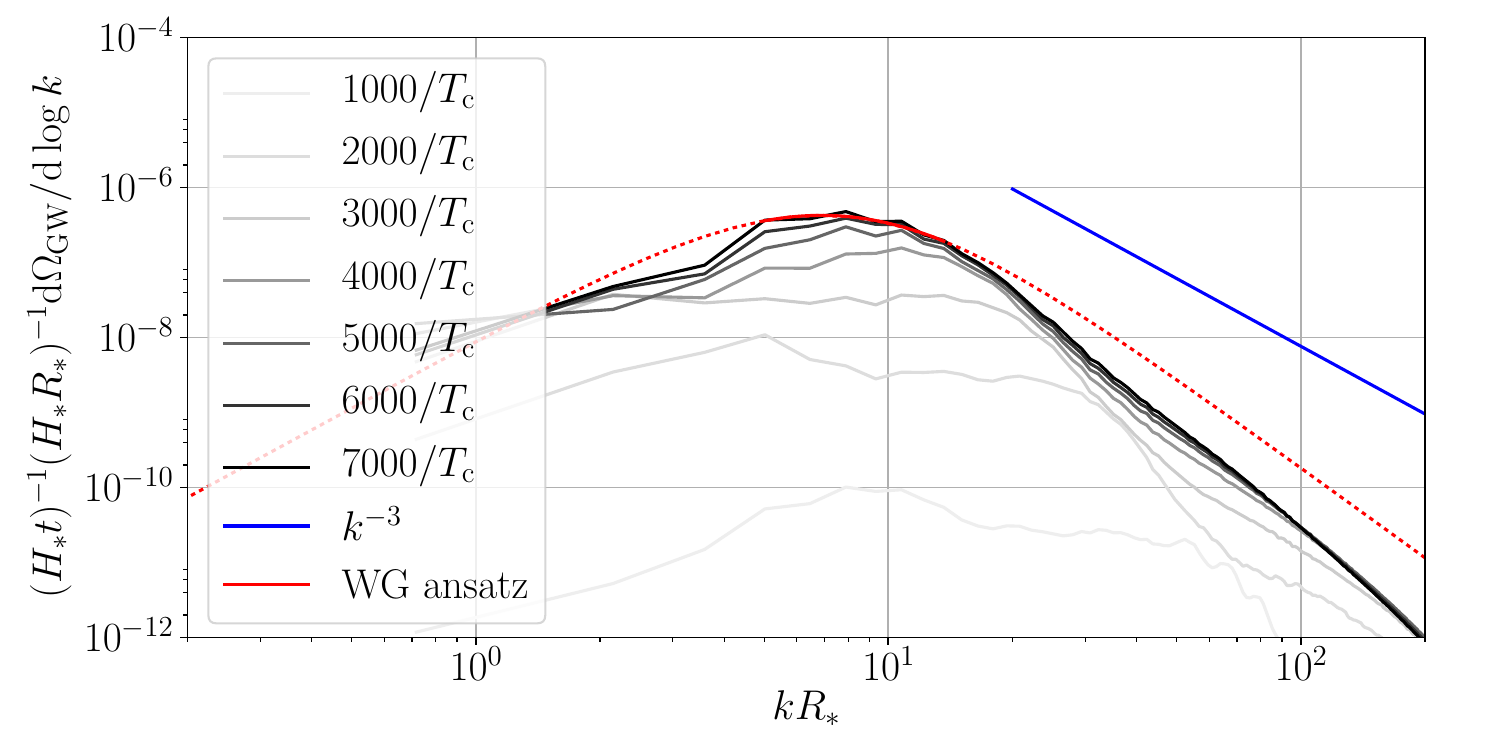}}
\end{figure*}

\begin{figure}
\flushleft
\subfigure[\ $\Nb=11$, $dx=1$]{\includegraphics[width=0.49\textwidth,clip=true]{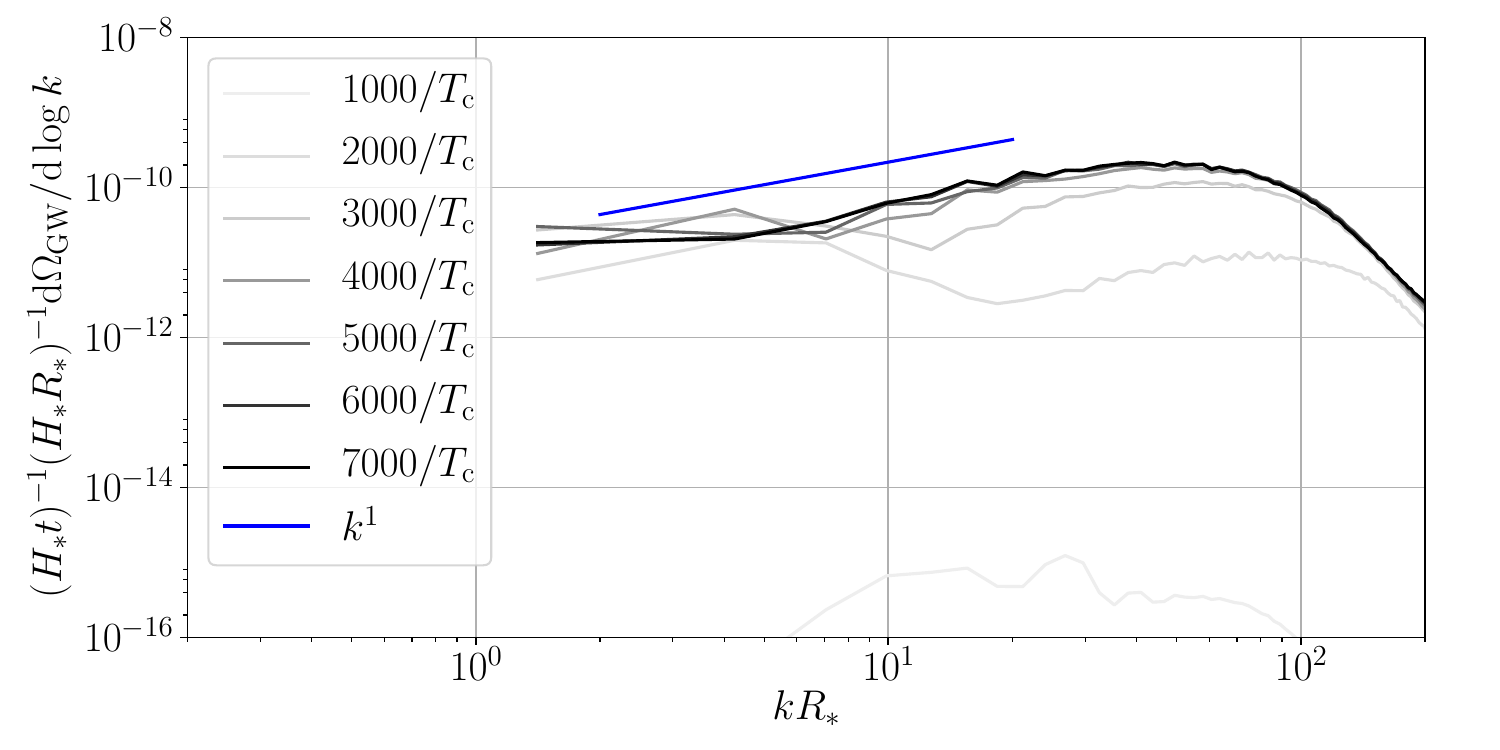}}
\hfill	
\\
\flushleft
\subfigure[\ $\Nb=84$]{\includegraphics[width=0.49\textwidth,clip=true]{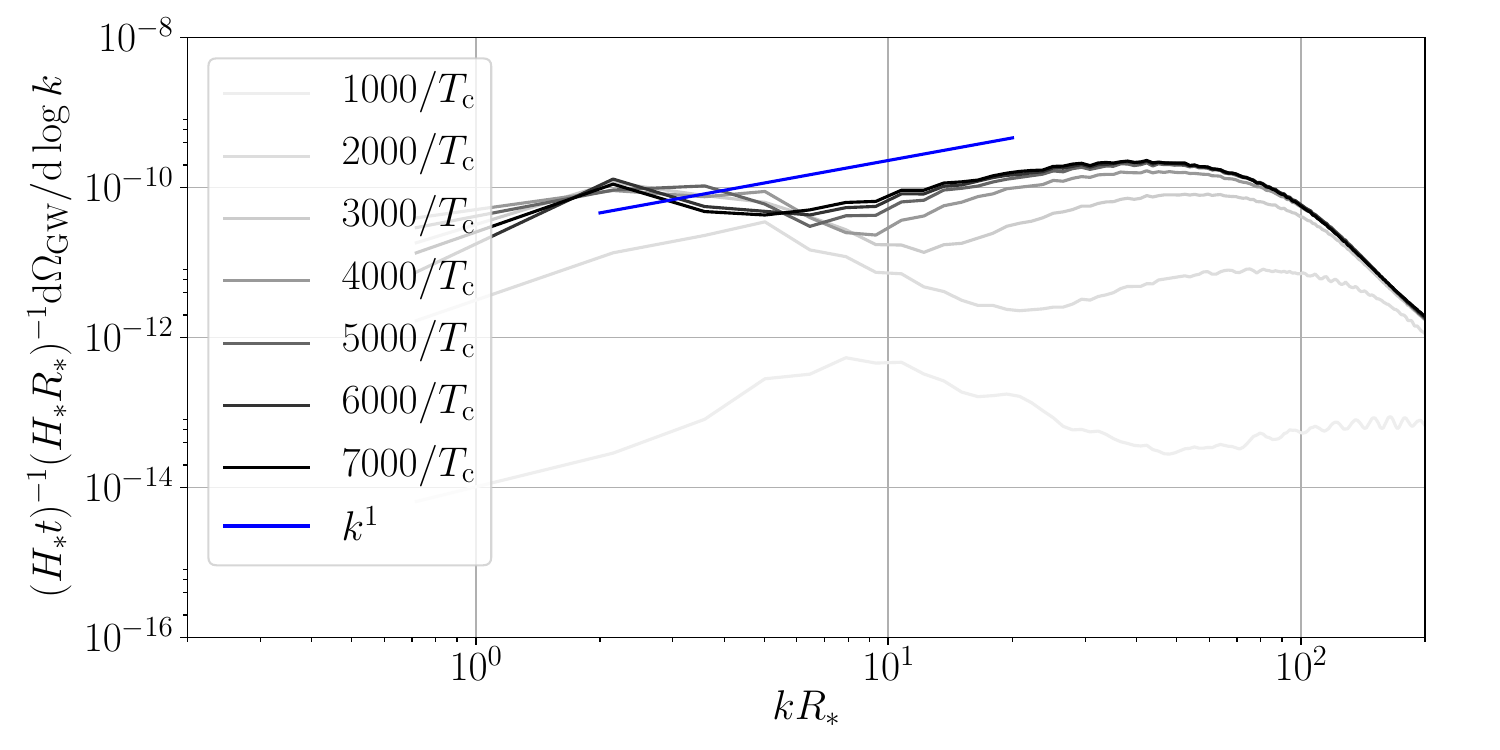}}
\hfill	
\\
\flushleft
\subfigure[\ $\Nb=5376$]{\includegraphics[width=0.49\textwidth,clip=true]{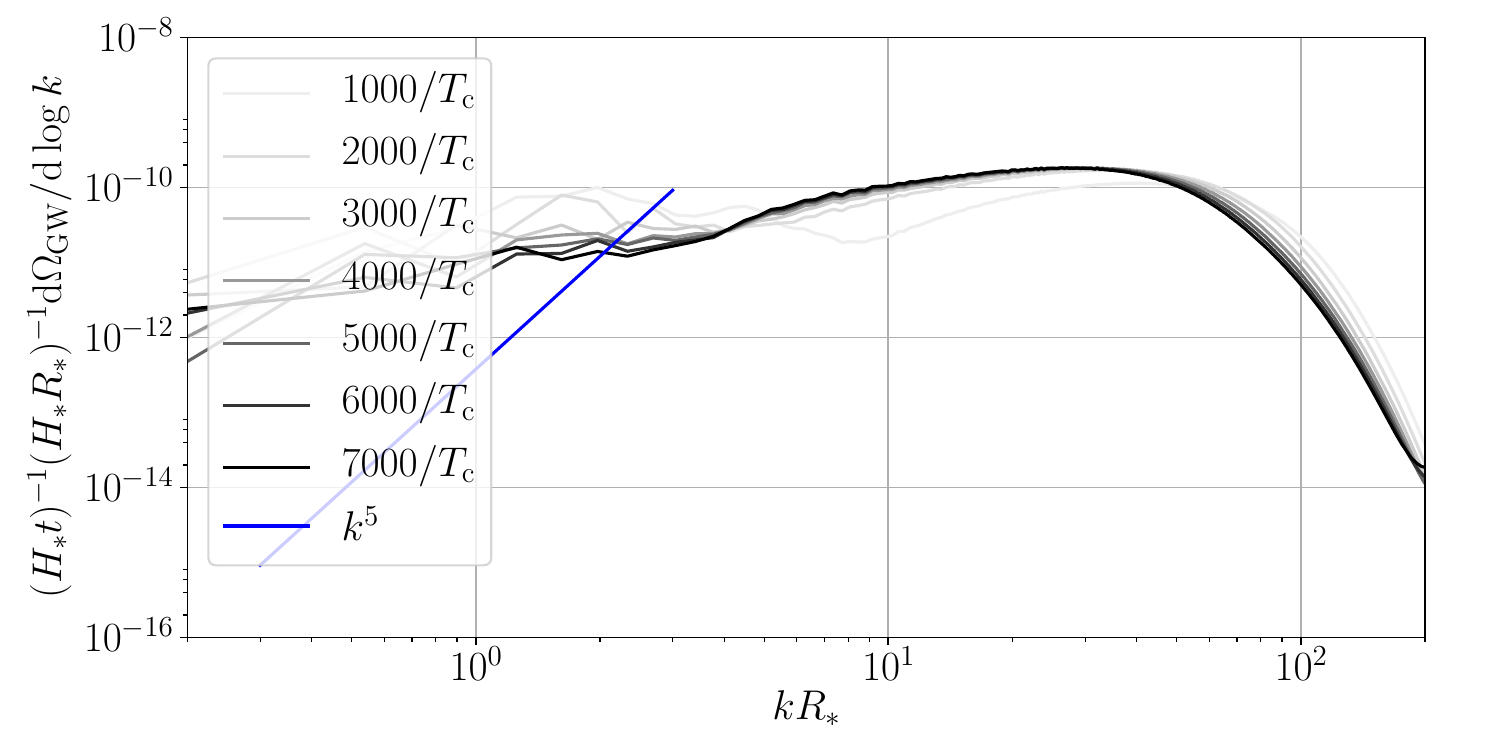}}
\hfill

\caption{\label{f:GWPSDefJou} Power spectra of fractional energy
  density in gravitational waves for near-Jouguet deflagrations, $\vw
  = 0.56$, divided by the ratio of the mean bubble separation $\Rbc$
  to the Hubble length at the transition $\HN$, and the ratio of the
  time to the Hubble time.  The wave number is scaled by the mean
  bubble separation.  All are weak transitions with $N_b = 11, 84$ and
  $5376$ ($\Rbc = 1889/\Tc$, $1918/\Tc$, and $480/\Tc$).}
\end{figure}

In Figs.~\ref{f:GWPSDet} to \ref{f:GWPSDefJou} we plot power spectra
of the fractional energy density in gravitational waves.  The spectra
are divided by the mean bubble separation $\Rbc$ and the time $t$ in
units of the Hubble distance and time, and plotted against wavenumber
in units of the inverse bubble separation, for ease of comparison with
Eq.~(\ref{e:GWPowSpe}).  Taking the fluid flow length scale to be
$\Rbc$, we have
\begin{equation}
\frac{1}{(\HN t) (\HN\Rbc) }\frac{d \OmGW(k)}{d \ln(k)} =  3\AdInd^2 \fluidV^4  \PspecGW(k\Rbc).
\label{e:pspec}
\end{equation}
Note that by dividing by time, gravitational wave power generated in
the collision phase will decrease, while acoustic phase gravitational
wave power will asymptote to a constant.  Note also that, by dividing
by $\HN^2$, we arrive at a quantity which is independent of $G$.

One can see that, at late times, the shape of the power spectrum
appears to change little, and is settling down to a characteristic
shape.  We would expect power laws to be less clear in the
gravitational wave power spectrum than in the velocity power spectrum,
as the former is a convolution of the latter over a range $\De k = \pm
\cs k$ at wavenumber $k$ \cite{Hindmarsh:2015qta}.

Where power laws are established over a sufficient range, we expect
that a velocity power spectrum going as $k^{n}$ should produce a
gravitational wave power spectrum going as $k^{2n-1}$. The sound shell
model \cite{Hindmarsh:2016lnk} predicts $n=-1$ for $\Rbc/\De\Rbc \ll
k\Rbc \ll \Rbc/\CorLen$, $n=1$ for $ 1 \ll k\Rbc \ll \Rbc/\De\Rbc $
(if the scales $\Rbc$ and $\De\Rbc $ are well separated), and $n=3$
for $ k\Rbc \ll 1$.

In Fig.~\ref{f:GWPSDet} we show the gravitational wave power spectra
from detonations ($\vw = 0.92$, $0.80$, and $\vw \simeq 0.7$) for
transitions of weak and intermediate strength, arising from the
velocity fields with power spectra in Fig.~\ref{f:VelPSDet}.  Again,
the general form is a broken power law, with a domed peak at $k\Rbc =
\textrm{O}(10)$.  As with the velocity power spectrum, the shape is
similar between the weak and intermediate cases at the same wall
velocity; the intermediate strength transitions have higher amplitude,
resulting from the higher RMS velocity.  We see that the broader dome
in the velocity power spectrum translates to a broader feature in the
gravitational wave spectrum, although the features around the peak are
not well resolved, particularly in the intermediate strength
transition.

The power-law to the right of the dome is close to the $k^{-3}$
predicted by the sound shell model, particularly in the the
detonations.  The predicted $k^{5}$ power law at long-wavelengths is
difficult to discern, as it is buried under a feature established
during the collision phase.

In Fig.~\ref{f:GWPSDef} we show gravitational wave power spectra from
a deflagrations with $\vw=0.44$, for weak and intermediate strength
transitions.  The power law at high $k$ is steeper than $k^{-3}$,
inconsistent with the sound shell model, but consistent with velocity
power spectra steeper than $k^{-1}$.

In Fig.~\ref{f:GWPSDefJou} we show the gravitational wave power
spectra from a deflagration with speed $\vw = 0.56$, where the sound
shell is very thin, and there are clearly two scales in the power
spectra.  Again, the low-$k$ behaviour is hidden behind the
gravitational waves from the collision phase due to the limited
duration of the simulation, although there is a suggestion of a
steepening below $k\Rbc \sim 5 $ in the case with the maximum
long-wavelength resolution, $\Nb = 5376$.  It is clear that the peak
is at around $k\Rbc \simeq 50$, which is understandable in terms of
the scale $k\De\Rbc \simeq 2$.

\section{Modelling the power spectrum for LISA}
\label{sec:lisa}

Our simulations nucleate all bubbles simultaneously, whereas in a real
thermal phase transition away from metastability, the nucleation rate
rises exponentially as $p(t) = p_0 \exp[\be(t - \tf)]$ after the
temperature drops below the critical temperature, where $\be$ is the
transition rate parameter, and $\tf$ is the time at which the volume
fraction of the symmetric phase is $1/e$ \cite{Enqvist:1991xw}.  This
means that there are a few larger, bubbles earlier in the transitions,
and more smaller bubbles as the transition ends.  Numerical
experiments with the gravitational wave power spectrum in the envelope
approximation \cite{Weir:2016tov} show that the principal effect of
instantaneous nucleation for a given mean bubble separation is to
increase the peak frequency by a factor of approximately 1.7, and
decrease the amplitude by a factor of about 3. The shape of the power
spectrum is not significantly changed.  This rescaling can be regarded
as a rescaling of the relationship between the mean bubble separation
$\Rbc$ and the transition rate parameter $\be$.

We shall assume that the same is true for the velocity field generated
by the colliding bubbles, i.e.~that the shape is not significantly
changed by using a realistic nucleation history, and that the
principal effect is to change the proportionality constant in the
standard relation $\Rbc = (8\pi)^{\frac{1}{3}} \vw/\beta$
\cite{Enqvist:1991xw}.  With this assumption we can directly use our
measured power spectra as models for the gravitational wave power
spectrum from a phase transition.

In the LISA Cosmology Working Group report \cite{Caprini:2015zlo} the
acoustic gravitational wave power spectrum was modelled using a broken
power law function:
\begin{equation}
 \frac{d \OmGW(k)}{d \ln(k)} 
=  (\HN\Rbc) A \fitfun(s)
\label{e:WGfit}
\end{equation}
where
\begin{equation}
\label{e:FitFun}
\fitfun(s) = s^3\left(\frac{7}{4+3s^2}\right)^{7/2},
\end{equation}
and 
\begin{equation}
\label{eq:ratio}
s = \frac{k\Rbc}{(k\Rbc)_{\rm max}}.
\end{equation}
The dimensionless parameters $A$ and $(k\Rbc)_{\rm max}$ determine the
magnitude and the location of the maximum of the power spectrum,
respectively.  The form of the function is motivated by the results
from hydrodynamical gravitational wave production simulations in
Ref.~\cite{Hindmarsh:2015qta}.  The power spectrum of the ansatz at
small $k$ is $\propto k^3$, turning over to $\propto k^{-4}$ at large
$k$.

The detailed structure of the ansatz determines the width of the dome
between the small-$k$ and large-$k$ regions; indeed, the width of the
dome can be adjusted by generalising the ansatz to a form
\begin{equation}
   \fitfun_a(s) =   s^3\left(\frac{7}{4+3s^a}\right)^{7/a}.
\end{equation}
In order to enable direct comparison with the working group results we
fit the measured power spectrum using the original ansatz
(\ref{e:WGfit}).

The resulting fits are shown as dashed lines in Fig.~\ref{f:GWPSDet}
for detonations and in Fig.~\ref{f:GWPSDef} for deflagrations, and fit
parameters are listed in Table \ref{t:WGfit}.  The functions are
fitted only in the neighbourhood of the domes, but at least for the
weak detonations the fitted functions describe the overall behaviour
of the data surprisingly well.

By integrating (\ref{e:WGfit}) and comparing to
Eq.~(\ref{e:GWTotPow}), we can derive an estimate for the amplitude
parameter from our simulations\footnote{There is a missing factor of 3 in this expression, see the erratum at the end of this paper.},
\begin{equation}
\label{e:Aest}
A_\text{est} \simeq  0.687 \Ga^2 \fluidV^4 \OmGWscaled^{\Rbc}.
\end{equation}
The estimates, computed from the values of $\fluidV$ in Table
\ref{t:SimVelStats} and $\OmGWscaled^{\Rbc}$ in Table \ref{t:SimGWs},
are shown in the last column of Table \ref{t:WGfit}.

\begin{table}
\begin{tabular}{lccccc}
Type & $\vw$ & $\Rbc$ & $(kR_*)_{\rm max}$ & $A$ & $A_\text{est}$   \\
\hline
Weak 
 & 0.92 & 1918 &  ~8.6 & $1.4\times10^{-11}$ & $1.9\times10^{-11}$  \\ 
 & 0.80 & 1918 &  10.4 & $3.1\times10^{-11}$ & $5.9\times10^{-11}$ \\ 
 & 0.68 & 1918 &  18.3 & $8.1\times10^{-11}$ & $14\times10^{-11}$ \\ 
 & 0.44 & 1918 &  ~9.9 & $8.2\times10^{-11}$ & $12\times10^{-11}$ \\ 
\hline
Int. 
 & 0.92 & 1918 & ~8.5 & $1.6\times10^{-7}$ & $2.7\times10^{-7}$ \\
 & 0.72 & 1889 & 16.1 & $3.7\times10^{-7}$ & $13\times10^{-7}$ \\ 
 & 0.44 & 1918 & ~6.9 & $4.3\times10^{-7}$ & $5.3\times10^{-7}$ \\ 
\hline
\end{tabular}
\caption{\label{t:WGfit} The fit parameters of the ansatz
  (\ref{e:WGfit}).  $(kR_*)_{\rm max}$ is the location of the maximum
  of the power spectrum, and $A$ its amplitude, along with the
  estimate from Eq. (\ref{e:Aest}).}
\end{table}

It can be seen that the amplitude estimates based on numerical
integration of the scaled gravitational wave power spectrum $\PspecGW$
are generally higher than those derived from the fit: this is because
the numerical power spectrum exaggerates the low-$k$ part of the power
spectrum derived from the collision phase. In the case of the
intermediate strength transition at $\vw=0.72$, the dome is less
apparent, and the fitting formula under-predicts the gravitational
wave spectrum at high $k$.  We recommend using the fitting formula for
wall velocities away from the Chapman-Jouguet speed by about $|\vw -
\vCJ| \gtrsim 0.1$.

For the near-Jouguet deflagration ($\vw=0.56$) the dome in the power
spectrum becomes very broad, as can be seen in
Fig.~\ref{f:GWPSDefJou}.  In this case the fit ansatz (\ref{e:WGfit})
cannot describe the behaviour well.  It is possible to construct more
complicated fit functions which can capture the structure at
intermediate scales, but the limited numerical data makes it difficult
to see universal features.  We leave the detailed analysis of the
Jouguet case for further analysis.

The peak angular frequency in units of the mean bubble separation $\zp
= (kR_*)_{\rm max}$ is generally around 10, except near the
Chapman-Jouguet speed where it is larger.  Qualitatively this agrees
with the estimate for the peak frequency made in
Ref.~\cite{Caprini:2015zlo} $\fp \simeq 1.2 \beta/\vw$, with the phase
transition rate parameter $\beta \simeq 3 \vw/\Rbc$.  For a more
precise estimate we must take into account the fact that nucleating
bubbles simultaneously changes the effective transition rate for a
given $\Rbc$ \cite{Weir:2016tov}, as outlined above.  A power spectrum
peaking at $(kR_*) = \zp$ in our simulations corresponds to a true
peak frequency
\begin{equation}
\label{e:fpThen}
\fp \simeq  \frac{0.54}{S} \frac{\beta}{\vw} \frac{\zp}{10}.
\end{equation}
where $S \simeq 2$ is a factor which takes into account the
overestimate of the frequency for a given $\Rbc$.

In order to calculate the observed frequency of waves emitted with
frequency $\fp$ (\ref{e:fpThen}) at time $\tN$ when the Hubble
parameter was $\HN$, we note that the peak frequency today can be
written
\begin{equation}
f_{\text{p},0} =  \frac{\fp}{\HN} H_{\text{n},0},
\end{equation}
Here 
\begin{equation}
H_{\text{n},0} = 16.5   \left(  \frac{\TN}{10^2 \, \text{GeV}} \right) \left(  \frac{h_*}{100}  \right)^{\frac{1}{6}} \; \mu\text{Hz}
\end{equation}
is the Hubble rate at the nucleation temperature, redshifted to today,
assuming that the dominant source of energy density is radiation.
Hence
\begin{equation}
f_{\text{p},0}
\simeq 26 \left( \frac{1}{\HN\Rbc} \right) \left( \frac{\zp}{10} \right) 
\left(  \frac{\TN}{10^2 \, \text{GeV}} \right) \left(  \frac{h_*}{100}  \right)^{\frac{1}{6}} \; \mu\text{Hz},
\end{equation}

To obtain the amplitude of the gravitational wave power spectra today,
the power spectrum (\ref{e:GWPowSpe}) and total power
(\ref{e:GWTotPow}) must be multiplied by a factor $F_{\text{gw},0} =
\Om_{\ga,0} (h_0/h_*)^{\frac{1}{3}}(h_0/2)$, where $\Om_{\ga,0}$ is
the density parameter of photons today, $h_0$ is the effective number
of relativistic degrees of freedom contributing to the entropy today,
and $h_* \simeq g_*$ is the corresponding number at the time of
gravitational wave generation.

Using the Planck best-fit value $H_0 = 67.8 \pm 0.9
\,\textrm{km}\,\textrm{s}^{-1}\,\textrm{Mpc}^{-1}$ \cite{Ade:2015xua},
and the Far Infrared Absolute Spectrophotometer (FIRAS) temperature
for the Cosmic Microwave Background $T_{\gamma,0} = 2.725 \pm 0.002\,
\textrm{K}$ \cite{Mather:1998gm}, we have
\begin{equation}
F_{\text{gw},0} = (3.57 \pm 0.05)\times 10^{-5}  \left(\frac{100}{h_*}\right)^{\frac{1}{3}}.
\end{equation}
Our final expression for the acoustic gravitational wave power spectrum today is\footnote{The numerical prefactor in this expression contains some errors; see the erratum at the end of this paper.}
\begin{equation}
\label{eq:final}
\frac{d \OmGWnow}{d \ln(f)} = 0.68 F_{\text{gw},0} \Ga^2 \fluidV^4 (\HN\Rbc) \OmGWscaled \fitfun\left(\frac{f}{\fpnow}\right).
\end{equation}

\begin{figure} 
   \centering
   \includegraphics[width=0.48\textwidth]{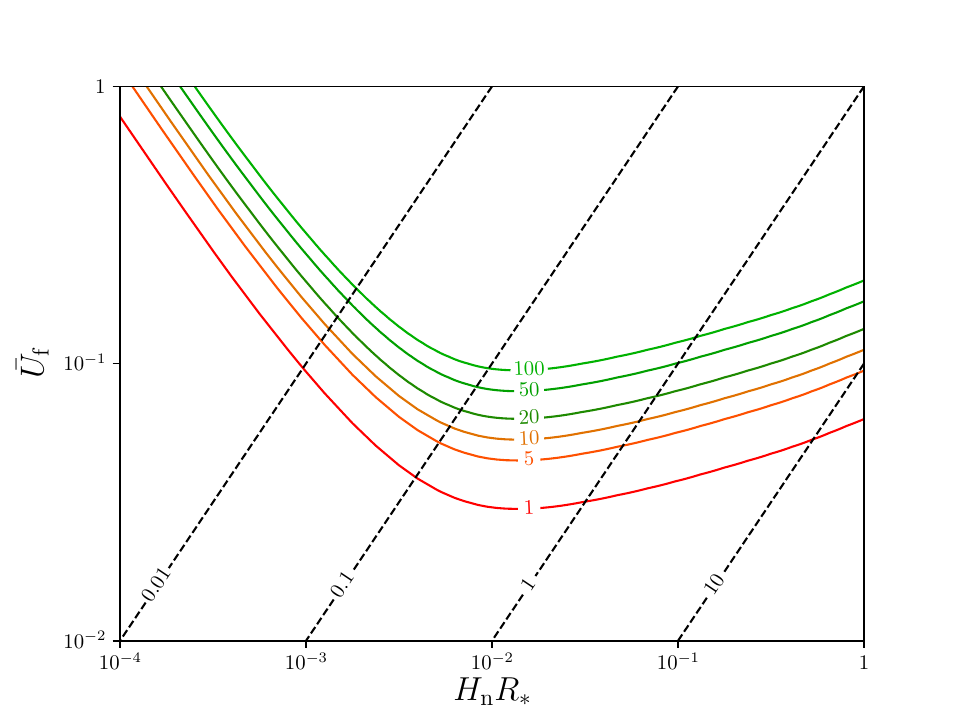} 
   \caption{Signal-to-noise ratio (solid contours) for the model power
     spectrum (\ref{e:WGfit}), using our acoustic peak gravitational
     wave power estimate (\ref{e:Aest}), with mission configuration
     described in the text, and taking the nucleation temperature to
     be $\TN = 100\;\textrm{GeV}$.  Dashed lines show the ratio of the
     order of magnitude of the shock appearance timescale
     $\Rbc\HN/\fluidV$.  For ratios of order 1 the fluid flow can be
     assumed to be purely acoustic, while for values much less than 1,
     the fluid flow may become turbulent within a Hubble time, and
     further investigation is required.\footnote{This plot is updated in the erratum at the end of this paper.}  }
   \label{f:contour}
\end{figure}

In Fig.~\ref{f:contour} we show the signal-to-noise ratio expected at
a LISA-like gravitational wave observatory with 6 laser links of arm
length $2$ Gm and a 5-year mission duration \cite{Klein:2015hvg}
(intermediate between configurations C1 and C2 of the LISA Cosmology
Working Group report \cite{Caprini:2015zlo}).  The SNR is shown as
contours in the $(\fluidV,\HN\Rbc)$ plane, for a phase transition with
a reference temperature of $\TN = 100\, \textrm{GeV}$, and taking\footnote{This should read $\OmGWscaled = 1.2\times 10^{-2}$. See the erratum.}
$\OmGWscaled = 1.2\times 10^{-1}$.  This value reproduces the peak
amplitude for the intermediate strength transition with $\vw=0.92$. It
under-predicts the power spectrum for the intermediate strength
transition at other wall speeds we simulated, and is therefore a
conservative estimate.

We also show contours of $\Rbc\HN/\fluidV$, which gives the timescale
for the appearance of shocks relative to the Hubble time.  We do not
expect the acoustic gravitational wave power spectrum to be an
accurate description for $\Rbc\HN/\fluidV \ll 1$.

From the figure, we can conclude that, for a phase transition at 100
GeV, LISA's peak sensitivity will be at a mean bubble separation of
about a hundredth the Hubble length, at which scale an intermediate
strength deflagration or detonation ($\fluidV \simeq 5 \times
10^{-2}$) should produce a SNR of around 10, taken as the detection
threshold in \cite{Caprini:2015zlo}.  However, this is also in the
region where one cannot safely assume that the fluid has not developed
shocks.  Further simulations are required for a more accurate
determination of LISA's ability to explore the parameter space of
phase transitions.

\section{Conclusions}
\label{sec:conclusions}

We have performed the largest numerical simulations to date of first
order phase transitions in the early universe, and computed the
resulting fluid velocity and gravitational wave spectra for a range of
bubble wall speeds $\vw$, mean bubble separations $\Rbc$, and
transition strengths $\StrParA$.  The power spectra are more tightly
pinned down than in our previous campaign of simulations
\cite{Hindmarsh:2015qta}.

We observe a gravitational wave power spectrum with a rising power
law, a broad dome at $k\Rbc = \textrm{O}(10)$, and a decreasing power
law at higher $k$.  The dome widens to a slowly rising plateau for
bubble wall speeds close to the speed of sound.

In the case of detonations the spectra exhibit a $k^{-3}$ power law in
good agreement with the sound shell model \cite{Hindmarsh:2016lnk},
but the high-$k$ power law for deflagrations appears slightly steeper.

We establish that there are two length scales in the gravitational
wave power spectrum: besides the mean bubble separation $\Rbc$ there
is also the mean sound shell thickness $\De\Rbc = \Rbc\De \vw/\cs$.
For a thin sound shell, where $\De\Rbc \ll \Rbc$, the slowly rising
plateau has a power law consistent with the $k^1$ predicted by the
sound shell model.

We do not have sufficient computational volume to determine the
gravitational wave power spectrum at $k\Rbc \lesssim \textrm{O}(1)$,
but the velocity power spectra for generic wall speeds are consistent
with $k^3$ there, which should lead to $k^5$ in the gravitational
waves.  In the special case of a just-subsonic deflagration, where the
sound shell is very thin, the velocity power spectrum is steeper than
$k^3$ for $k\Rbc \lesssim \textrm{O}(1)$. This is not in contradiction
with the sound shell model, which only gives $k^3$ for $k\Rbc \lesssim
\textrm{O}(|\vw - \cs|)$.

Our simulations are still not large enough to properly resolve all the
different wavenumber regimes simultaneously, and they reveal an
interesting knee in the velocity power spectra for deflagrations at
high $k$.  This feature appears to be established at about the peak
bubble collision time $\tColl$, where the area of the phase boundary
is at its maximum.  Yet larger simulations are therefore needed to
resolve the full power spectrum, and a separate simulation campaign to
investigate the deflagration power spectra.

We also need to run the simulations for longer, with larger fluid
velocities, in order to investigate the transition to turbulence.
Flows with larger fluid velocities become turbulent earlier, and are
likely to be important for the gravitational wave signals observable
by LISA.

However, for transitions which are weak enough that turbulence does
not develop, and have bubble wall speeds not too close to the speed of
sound, we are confident in the form of the power spectrum, and offer
the fitting formula (\ref{e:WGfit}) for the gravitational wave power
spectrum.  This can safely be used for RMS fluid velocities up to
$\fluidV \lesssim \Rbc \HN$, or $\kappa_\text{v}\StrParB \lesssim
(\HN/\beta)^2$ in terms of the kinetic energy conversion efficiency
$\kappa_\text{v}$ and the transition strength parameter $\StrParB$
(see Eq.~(\ref{e:KapDef})).
  
Using the fitting formula, we computed the signal-to-noise ratio at a
LISA-like mission with 6 laser links, arm length 2 Gm, and a duration
of 5 years.  At a transition temperature of $\TN = 100\;\textrm{GeV}$,
we find that such a mission has greatest sensitivity to transitions
with mean bubble separation of about one hundredth of the Hubble
length.  At this bubble separation, LISA will be able to detect the
signal from phase transitions down to latent heat to energy ratio
$\StrParA \simeq 0.1$, i.e.~intermediate strength in our terminology.

We also checked the timescale for the development of shocks and
turbulence, finding that a significant fraction of the detectable
parameter space is in a region where one cannot safely assume that the
fluid flow remains linear.  Further simulations are required to
examine the transition to turbulence, necessary for a more accurate
determination of LISA's ability to explore the parameter space of
phase transitions.

There is a range of extensions of the standard model (SM), which could
produce a gravitational wave signal in reach of LISA
\cite{Caprini:2015zlo}. Most of these models are extensions of the SM
Higgs sector by additional scalar fields, principally SU(2) singlets
or doublets, or modifications of the SM Higgs potential itself.
Furthermore, LISA could also probe phase transitions in the TeV range,
e.g. possible confinement transitions in strong coupling completions
of the electroweak theory.

A strong phase transition in these models can either come from 1) a
modified zero temperature Higgs potential, as is often the case in
singlet extensions, or 2) additional thermal contributions to the
Higgs potential, or 3) a combination of both, as e.g. in the
Two-Higgs-Doublet model (2HDM). Models of type 2) and 3) will show a
stronger dependence of the nucleation rate on temperature (via the
energy of the critical bubble), which results in a smaller value of
$\beta/H$. So models of type 1) will have larger bubbles at fixed
$\alpha$ compared to types 2) and 3).  As a result models of type 1)
may show a stronger gravitational wave signal without reaching the
state of turbulence.

Most of the models will show an observable gravitational wave signal
signal only when the walls propagate as fast detonations, but there
may be exceptions as has been shown e.g. for the 2HDM
\cite{Dorsch:2016nrg}.

More generally, the fact that the shape of the gravitational wave
power spectrum depends on the bubble wall speed means that it could
become possible to determine $\vw$ from a detection of the stochastic
gravitational wave background with sufficiently high signal-to-noise
ratio.

Delineating the ability of future space-based gravitational wave
detectors to constrain the full set of phase transition parameters
$\al$, $\be$, $\vw$ and $\TN$ of an electroweak-scale first order
transition, thereby opening a new quantitative window onto physics
beyond the Standard Model, is now an important goal.

\begin{acknowledgments}

The LISA Cosmology Working Group has provided an important forum for
discussing our work and we thank its members as well as its
coordinators, Chiara Caprini and Germano Nardini. We owe a huge debt
also to the late Pierre Bin\'etruy for his work on cosmological
sources of gravitational waves, his work with LISA, and in particular
his encouragement of our work.

We acknowledge PRACE for awarding us access to resource HAZEL HEN
based in Germany at the High Performance Computing Center Stuttgart
(HLRS). Our work was supported by the Munich
Institute for Astro- and Particle Physics (MIAPP) of the DFG cluster
of excellence ``Origin and Structure of the Universe''.

MH (ORCID ID 0000-0002-9307-437X) and SH 
acknowledge support from the Science and Technology Facilities
Council (grant number ST/L000504/1).  KR (ORCID ID
0000-0003-2266-4716) is supported by the Academy of Finland grant
267286. DJW (ORCID ID 0000-0001-6986-0517) was supported by the People
Programme (Marie Sk{\l}odowska-Curie actions) of the European Union
Seventh Framework Programme (FP7/2007-2013) under grant agreement
number PIEF-GA-2013-629425. 
\end{acknowledgments}

\bibliography{prace-results-paper}

\newpage

\begin{widetext}

\section*{Erratum}

\setcounter{equation}{0}
%\setcounter{figure}{0}

% In Eq.~\ref{e:WGfit}, the ratio $s$ on the right hand side implicitly depends on the same $k$ as the left hand side (through Eq.~\ref{eq:ratio}). This may not be immediately clear from the notation used.
%
% \bigskip
  
There is a factor of 3 missing from the right hand side of Eq.~\ref{e:Aest}, which should read:
\begin{equation}
  A_\text{est} \simeq  2.061 \Ga^2 \fluidV^4 \OmGWscaled^{\Rbc}.
\end{equation}
Without this additional factor of 3 this expression is not consistent with Eqs.~\ref{e:GWPowSpe} and~\ref{e:GWTotPow}, nor with Ref.~\cite{Hindmarsh:2015qta}. However, the values of $A_\text{est}$ in Table~\ref{t:WGfit} \textsl{are} correct and include this factor.

This factor of 3 is, in turn, missing from Eq.~\ref{eq:final}, which should read (note that original equation also truncated the numerical coefficient as 0.68 rather than 0.687):
\begin{equation}
\label{eq:corrected}
  \frac{d \OmGWnow}{d \ln(f)} = 2.061 F_{\text{gw},0} \Ga^2 \fluidV^4 (\HN\Rbc)
\OmGWscaled \fitfun\left(\frac{f}{\fpnow}\right).
\end{equation}
We also wish to highlight that this formula is written without the customary factors of the reduced Hubble constant $h^2$ on each side. This itself does not affect the validity of the equation and indeed allows us to select a value of $h$ for comparison with a specific precomputed gravitational wave energy density sensitivity curve.

Note also that in the paragraph below Eq.~\ref{eq:final} the exponent in the value for $\tilde{\Omega}_\text{gw}$ is incorrect. The expression should read $\tilde{\Omega}_\text{gw} = 1.2\times 10^{-2}$.

The SNR contours in Fig.~\ref{f:contour} were produced using
Eq.~\ref{eq:final}. The above factor of 3 (mistakenly absent from
Eq.~\ref{eq:final}) was \textsl{included} but the factor of 0.68[7]
(present in Eq.~\ref{eq:final}) was \textsl{excluded}, and there was
an additional erroneous factor of 2. This omission means the correct
predicted SNR is a factor of $0.687\times 0.5 \approx 0.34$ lower than
in the original plot.

Figure~\ref{fig:erratum} below is a corrected version of
Fig.~\ref{f:contour} that takes account of the numerical errors
explained above.

\renewcommand{\thefigure}{1}
\begin{figure*}[h!]
  \includegraphics[width=0.4\textwidth]{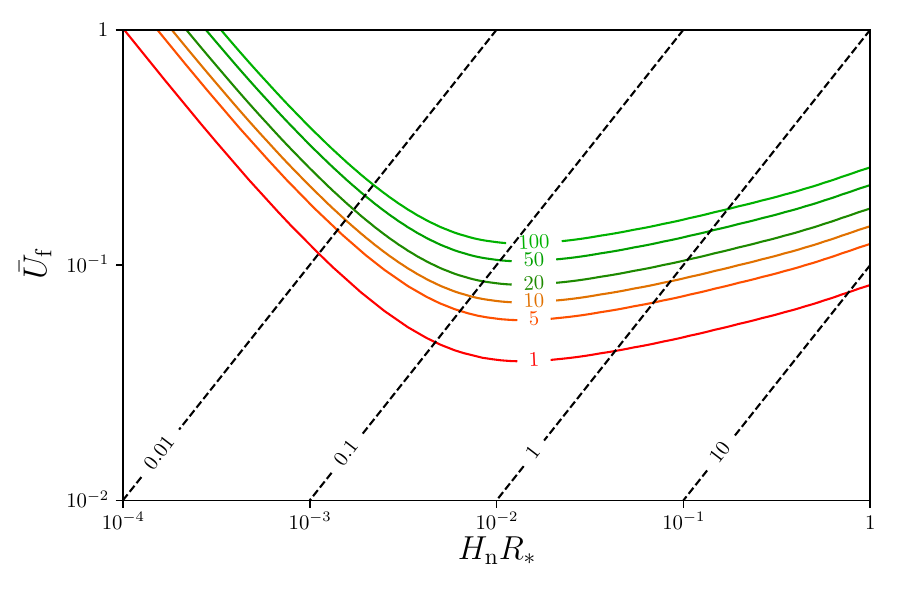}

\caption{\label{fig:erratum} Corrected version of
  Fig.~\ref{f:contour}, corresponding to Eq.~\ref{eq:corrected}
  above. The missing factor of 3 in Eq.~\ref{eq:final} was already
  included, but the 0.687 numerical prefactor was
  missing. Furthermore, an additional factor of 2 was mistakenly
  present. The overall effect was to reduce the SNR for a given
  parameter choice by a factor of approximately 0.34 compared to the
  original figure. Note that the mission profile for LISA used here is
  no longer current, so this plot is for qualitative comparison with
  Fig.~\ref{f:contour} only.}

\end{figure*}

However, the mission profile of LISA has changed substantially since
this paper was published, with a longer expected mission duration and
a modified sensitivity curve. This will further change the predicted
SNR curves. The plots in the following referenced paper are free of
the errors discussed in the erratum, and use the most up-to-date
mission profile\footnote{The parameter space can also be explored at
  \href{http://www.ptplot.org/ptplot/}{ptplot.org}.}.

\acknowledgments

We are grateful to Aleksandr Azatov, Francesco Sgarlata and Daniele Barducci for bringing this to our attention.

\section*{Reference}

\noindent C.~Caprini {\it et al.},
  ``Detecting gravitational waves from cosmological phase transitions with LISA: an update,''
\href{http://dx.doi.org/10.1088/1475-7516/2020/03/024}{JCAP \textbf{2003}, 024 (2020)}, 
\href{http://arxiv.org/abs/1910.13125}{[arXiv:1910.13125 [astro-ph.CO]]}.

\end{widetext}
  
\end{document}